\documentclass[preprint,12pt,authoryear]{elsarticle}

\usepackage{amssymb}
\usepackage{amsthm}
\usepackage{amsmath}

\usepackage{hyperref}

\usepackage{graphicx}
\graphicspath{ {./figures/} }

\usepackage{booktabs}
\usepackage{tabularx}
\newcolumntype{C}[1]{>{\centering\arraybackslash}p{#1}}
\newcolumntype{L}[1]{>{\raggedright\arraybackslash}p{#1}}
\newcolumntype{R}[1]{>{\raggedleft\arraybackslash}p{#1}}

\usepackage{color}
\usepackage{upgreek}

\usepackage[table,xcdraw]{xcolor}

\journal{Treatise on Geochemistry}

\begin{document}

\begin{frontmatter}



\title{Venus}


\author[inst1]{Cedric Gillmann}
\affiliation[inst1]{organization={ETH Zürich, Institute of Geophysics}, addressline={Sonneggstrasse 5, NO H 25}, city={Zürich}, postcode={8092}, country={Switzerland}}

\author[inst11]{Giada N. Arney}
\affiliation[inst11]{organization={NASA Goddard Space Flight Center}, city={Greenbelt},postcode={20771}, country={United States}}

\author[inst6]{Guillaume Avice}
\affiliation[inst6]{organization={Université Paris Cité, Institut de physique du globe de Paris, CNRS}, addressline={1, rue Jussieu}, city={Paris},postcode={75005}, country={France}}

\author[inst7,inst8]{M. D. Dyar}
\affiliation[inst7]{organization={Planetary Science Institute}, addressline={1700 East Fort Lowell, Suite 106}, city={Tucson},postcode={AZ 85719}, country={United States}}
\affiliation[inst8]{organization={Mount Holyoke College}, addressline={50 College St.}, city={South Hadley},postcode={MA 01075}, country={United States}}

\author[inst2]{Gregor J. Golabek}
\affiliation[inst2]{organization={Bayerisches Geoinstitut, University of Bayreuth}, addressline={Universitätsstrasse 30}, city={Bayreuth},postcode={95440}, country={Germany}}

\author[inst3,inst4]{Anna J. P. G\"{u}lcher}
\affiliation[inst3]{organization={Jet Propulsion Laboratory, California Institute of Technology}, addressline={4800 Oak Grove Drive}, city={Pasadena},postcode={CA 91109}, country={United States}}
\affiliation[inst4]{organization={Division of Geology and Planetary Sciences, California Institute of Technology}, addressline={1200 E California Blvd}, city={Pasadena},postcode={CA 91125}, country={United States}}

\author[inst11]{Natasha M. Johnson}

\author[inst5]{Maxence Lef\`evre}
\affiliation[inst5]{organization={LATMOS/IPSL, Sorbonne Université, UVSQ Université Paris-Saclay, CNRS}, addressline={4 Place Jussieu}, city={Paris},postcode={75252 Cedex 05}, country={France}}

\author[inst9,inst10]{Thomas Widemann}
\affiliation[inst9]{organization={LESIA, Observatoire de Paris, Université Paris Sciences et Lettres}, city={Meudon},postcode={92195}, country={France}}
\affiliation[inst10]{organization={UVSQ, Université Paris-Saclay}, city={Guyancourt},postcode={78047}, country={France}}



\begin{abstract}
After decades of relative neglect, interest in Venus surges anew in the planetary science community and the public. New missions are planned and selected, and will pave the way to the decade of Venus, as new observations allow us to uncover some of the many mysteries our closest Solar System neighbor still harbors. Building on the legacy of past works, here, we discuss the state of our understanding of Venus, from both observation and modeling. We describe each of the envelopes of the planet, from its atmosphere to its interior with an eye for the most recent advances and current topics of interest. We then briefly discuss coupled modelling efforts to better constrain the evolution of the planet. Finally, we describe how the upcoming missions and concepts will further lift the veil on Venus' secrets.
\end{abstract}

\begin{highlights}
\item We review the current state of Venus science.
\item Recent modeling work about Venus is summarized.
\item The current main hypothetical scenarios for the evolution of Venus are briefly discussed.
\item Upcoming missions to Venus and their expected benefits are described.
\end{highlights}


\end{frontmatter}


\section{Introduction}
\label{sec:sample1}

Venus has subverted the expectations of generations of scientists and teased stargazing humans for millennia. Yet, we are still very far from understanding our closest neighbour in the solar system. This chapter proposes an introduction to the current state of our knowledge. We did not try to gather here the alpha and omega to all things related to Venus, as other recent publications have discussed specific topics in much more details, and will be referred to as appropriate in the present contribution. Instead, we aim to build a reference of measurements that have been gathered until now, and how they are currently used to constrain plausible evolution scenarios of the planet up to its present-day state. We also briefly discuss the efforts that are currently deployed to fill in gaps in our knowledge, through missions and models. Finally, we remind the reader that this work does not replace the Venus chapter in the previous edition of the treatise \citep{fegley2014venus}, but aims at exploring Venus science from a different direction.

\subsection{Ancient history}
As the brightest light in the evening and morning sky aside from the Moon, Venus was likely known and recognized by humans since prehistoric times. Texts and archaeological remains from ancient civilizations certainly imply it had a strong influence around the world under many names \citep{grinspoon1997}. Some of the earliest references to Venus can be found in the Venus tablet of Ammisaduqa (dated back to the first millennium BC; see \citet{huber1982,de2010new}), in Mesopotamia, as astronomical observations regarding the rise and set times and location over a period of 21 years during the second millennium BC were noted. Due to the planet's appearance in the sky as either the evening star or the morning star, it was sometimes regarded as two distinct stars, but some civilizations such as the Babylonians and the ancient Greek appear to have been aware that it was a single body. The peculiar cycle of appearance and disappearance of Venus' light in the sky, due to its position in the inner Solar System, was noted early in history and is most likely the cause for the planet's significance in most of the early civilizations' cosmologies. The planet is often associated with a dual pattern of death/rebirth (Maya, Sumer), where transformation comes from upheaval and is often linked to love and war, as in Mesopotamia or Ancient Rome. For example, Inanna, the Sumerian precursor of the Greek Aphrodite already combines all those aspects together, from passion and strife to the journey into the underworld \citep[][]{marcovich1996,hughes2019}. The Maya were also especially interested in that planet, possibly more so than even the moon, and built calendars predicting its position for thousands of years \citep{sokol2022stargazers,aveni1986maya}. The Dresden Codex, one of the last codices remaining from that civilization, features Venus prominently. For them, Venus was associated with the god Kukulcan (Quetzalcoatl), and the cycles of Venus to the god's resurrections. The temples of Uxmal seem to have been oriented and designed for astronomical observation of Venus \citep{lamb1980sun}.
Although the Romans named the planet Venus Lucifer, from Greek tradition (the Bringer of Light), they associated it with Venus, the goddess of love, which led to its current name \citep{ovid1998metamorphoses}. Interestingly enough, the pattern of destruction and creation can also be found there: the goddess Venus is married to Mars (war) and is at the origins of the events that precipitate the Trojan War \citep{pope2009iliad}, the destruction of the city by the Greeks and subsequent birth of Rome from the descendants of Trojan refugees \citep{virgil2004aeneid}.

\subsection{Exploration history}
Perhaps fittingly, the ebb and flow of scientific interest in the planet Venus seems to have followed a similar path in more recent times. The early observations of Venus with a telescope by Galileo Galilei in 1610 revealed the planet's phases and confirmed the heliocentric model. Later observations of the transit of Venus in front of the Sun provided estimates of the astronomical unit and led to the first suggestions of the existence of Venus' atmosphere \citep[see][]{grinspoon1997}. However, despite attention and repeated observation, little actual knowledge was gathered by optical observations due to the mostly featureless appearance of its cloud cover in visible light. 

Since the Magellan mission, launched in 1989, only two dedicated missions have reached Venus (Akatsuki from JAXA and Venus Express from ESA). Further discussion of the history of Venus exploration can be found in \citet{grinspoon1997,bonnet2013history,fegley2014venus,orourke2023}. The most recent developments since the last edition of this treatise \citep[][]{fegley2014venus} include the conclusion of the Venus Express mission in 2014, and since then about 20 additional flybys by missions en route to other targets (BepiColombo, Solar Orbiter and the Parker Solar Probe) in addition to Akatsuki's mission in Venus orbit.

The dearth of missions after the 1980s was accompanied by a decline in general interest in Venus science compared to other targets such as Mars. The reasons for this are probably a combination of multiple factors. Venus was a prime object of Soviet space exploration, and the dissolution of the Soviet Union in 1991 led to the cancellation of several projects. The USA were more involved in martian exploration, building on the success of the Viking missions. Mars went on to be the next hot topic: the search for water became an overarching goal for exploration in the next decades due to the clear connection to habitability and life \citep[][]{brack1996,gillmann2023}. The inhospitable surface of Venus made it a much less desirable target than Mars. While cold, Mars evidenced clear signs of the presence of liquid water in the past, and familiar-looking polar caps. The disappointment following the revelation that Venus did not turn out to be the humid world that had been envisioned during the early 20$^{th}$ century probably also played a part in the reduced interest in further research \citep[see][]{orourke2023}. Additionally, while Venus missions historically had a high success rate, the hostile conditions at the planet's surface made building landers capable of resisting its surface conditions a much costlier endeavour, for less obvious scientific returns. As a result, measurements in the last decades focused on Venus' atmosphere: the easiest part of the planet to observe. With a limited number of new missions and little data, new research was often based on decade-old measurements and had a hard time building the momentum necessary to motivate new projects.

\subsection{Current return to the spotlight}
However, in recent years, the attention of the planetary community has once again turned towards Venus. Awareness that comparative planetology may hold the keys to answering long-standing questions about planetary evolution is on the rise. And Venus, as Earth's neglected sister, is at the top of the list of candidates to provide those answers. Two main factors are contributing to this renewal of interest. 

First, Venus is the closest analog to Earth that can be directly studied. Today, it has an extremely inhospitable and hot climate, due to a massive CO$_2$ atmosphere containing very little water, in stark contrast to the mild conditions that have prevailed on Earth for hundreds of millions of years at least. While it is clear that Mars' surface has been partly shaped by water, its smaller mass and greater distance from the Sun clearly set it apart from Earth. Venus is a prime example of a planet that could have been similar to Earth, early on, but evolved in a radically different way \citep[see][]{Hamano2023}. Recent work on Venus has also highlighted the links between the planet and the question of habitability, with, for example, the discussion of the plausibility of water oceans in the distant, or not so distant, past of the planet \citep[][]{WayDelGenio2020}. 
In summary, Venus may hold critical answers to how a planet becomes or ceases to be habitable, one of the major interrogations of modern planetology.

The second important factor is the discovery of thousands of new exoplanets and candidates, and the realization that Venus may be the only analogue to exoplanets that we can study in great detail and might even get our hands on. Indeed, exoplanets in the Venus zone (close to their star) will be the easiest terrestrial planets to observe and characterize \citep[][]{kane2019,way2023}. Exoplanet observations will also never be able to yield as much data and precision as we can hope from a planet in our own solar system. At the time of writing of this review, the James Webb Space Telescope (JWST) has already been used to identify CO$_2$ in the atmosphere of exoplanet WASP-39 b \citep[see][]{jwst2023} and H$_2$O in the atmosphere of WASP-96 b \citep[][]{nikolov2022solar}, both hot gas giant planets. It also took a spectrum, albeit featureless, of short-period Earth-sized exoplanet LHS 475 b \citep[][]{lustig2023jwst}, and is used to characterize the atmospheres of the terrestrial planets in the TRAPPIST-1 system \citep[e.g.][]{greene2023,zieba2023}, further cementing the future key role of Venus as analog for exoplanetary studies.

In this chapter, we will describe how various measurements and models have helped us to characterize Venus, starting with general and orbital properties, and then moving on to the atmosphere, interior and surface of the planet. We will close this review with two sections, respectively discussing the evolution of the planet through modelling, and how the near future missions will address some of the lingering questions about Venus.

\section{General characteristics of Venus}

As we mentioned above, the striking differences between present-day Earth and Venus, especially in terms of surface conditions, are only highlighted by the fact that no other solar system body shares as many similarities with Earth.
Venus orbits in the inner Solar System, just outside the inner boundary of the habitable zone \citep[][]{kane2019}. Its mean distance from the Sun is of the same order as Earth's, at 108.21~$\times$~10$^6$~km, or 0.72~au, with an eccentricity of 0.0068 and an orbital inclination of 3.395~degrees \citep[see][]{lodders1998}. One revolution of Venus around the Sun takes 224.7~Earth days, while the length of day is 116.75 Earth days. The inclination of the equator of Venus relative to its orbit is very small (2.64~degrees) compared to Earth, but with the peculiarity that Venus' slow rotation is retrograde. While the cause for this characteristic is still unknown, suggestions have included an early impact \citet{mccord1968,alemi2006}, core-mantle friction \citep{goldreich1970,correia2001, correia2003a,correia2003b}, momentum exchanges between the solid planet and its massive atmosphere \citep[e.g.][]{macdonald1964,WayDelGenio2020} or even an early ocean \citep{green2019}.

Venus' radius and mass have been measured at 6051.8~km, with no polar flattening \citep{seidelmann2007}, and 4.8673~$\times$~10$^{24}$~kg, respectively, for a corresponding mean density of 5243~kg~m$^{-3}$ (5513~kg~m$^{-3}$ for Earth) and a surface gravity of 8.87~m/s$^2$. Those properties suggest that the interior structure of Venus and Earth could be remarkably similar on the first order. However, it is notable that no self-generated or remanent magnetic field has been detected on Venus \citep[][]{russell1980}, contrary to both Earth and Mars, respectively, which implies significant differences in core conditions. Given the sensitivity of the Pioneer Venus Orbiter magnetometer, any dipole field would be weaker than 10$^{-5}$ that of Earth \citep[][]{phillips1987}.

\section{The atmosphere of Venus}

\subsection{Composition}

\subsubsection{Abundances}
The surface of Venus, discussed later in this review, is not observable in visible light from Earth nor from close orbit due to the thick cloud layers supported by its dense atmosphere. The total mass of the atmosphere of Venus is estimated to $\approx$ 4.8~$\times$~10$^{20}$~kg (vs 5.1~$\times$~10$^{18}$~kg on Earth), which translates into a surface pressure of 92~bar, and a surface density of $\approx$ 65~kg~m$^{-3}$. The atmosphere of Venus is dominated by CO$_2$ (96.5\% by volume) and has a composition closer to Mars' than to Earth's (despite being 10,000 times more massive than Mars' atmosphere). The rest of the atmosphere (at ground level) is mostly N$_2$ (about 3.5\%) and all other species are minor components, making up to 0.01\% of the total atmosphere. The following have been measured: Sulfur Dioxide (SO$_2$) at 150~ppmv (22-42~km altitude), Argon (Ar) at 70~ppmv, Water (H$_2$O) at 30~ppmv (but with an even drier upper atmosphere), Carbon Monoxide (CO) at 17~ppmv, Helium (He) at 12~ppmv and Neon (Ne) at 7~ppmv \citep[][]{lodders1998,marcq2008}. As a result, the mean molecular weight of the atmosphere of Venus is 43.45, much higher than Earth's 28.97. With a resulting scale height of about 16~km, Venus' atmosphere is comparable to the other planets in the solar system that sustain an atmosphere (8.5~km for Earth, 18~km for Mars).

\subsubsection{Mass equivalents and comparison with Earth}
The mass of the individual main components of the atmosphere of Venus are also sometimes used as constraints in modelling efforts and to discriminate between hypothetical evolution scenarios. The bulk of the atmosphere contains 4.69~$\times$~10$^{20}$~kg CO$_2$ \citep[][]{fegley2014venus} and 11~$\times$~10$^{18}$~kg N$_2$ \citep[][]{johnson2015}. The mass of SO$_2$ in the atmosphere has been estimated at about 6~$\times$~10$^{16}$~kg \citep[][]{fegley2014venus}, even though the apparent decrease in SO$_2$ mixing ratio by a factor of $\approx$ 4 near the surface is not yet understood \citep[][]{bertaux1996,krasnopolsky2007}, and notable latitudinal variations have been reported \citep[][]{marcq2021,oschlisniok2021}. It appears that despite taking up only 3.5\% of the atmosphere on Venus, N$_2$ is nearly three times more abundant than in Earth's atmosphere (3.89~$\times$~10$^{18}$~kg) and twice the content of Earth's superficial envelopes including the crust \citep[][]{lodders2015}. Additionally, while water represents a tiny part of the atmosphere, it still amounts to about 10$^{16}$~kg \citep[][]{kasting1983loss, lecuyer2000}. It is also remarkable that the total mass of CO$_2$ in the atmosphere of Venus is comparable to that in Earth's atmosphere and crust \citep[e.g. ][]{donahue1983origin,lecuyer2000,hartmann2012terrestrial}; a similarity that could suggest that most of Venus' CO$_2$ inventory is now in the atmosphere \citep[e.g.][]{lammer2008atmospheric}, but the topic is still debated (see the discussion of Argon isotopic ratio further down). 

\subsubsection{A note on common use in the literature}
Recent observations from the Akatsuki mission have highlighted that large variations exist within the atmosphere of Venus, both vertically and horizontally, but also with time, including important species, such as water, SO$_2$ and CO \citep[][]{marcq2018}. Given the mass of the Venusian atmosphere, it is not surprising that those variations can be much larger than on Earth.

Note that atmosphere compositions are usually given in volume fractions (mixing ratio), relating directly to partial pressures (the partial pressure is the total pressure times the volume mixing ratio), and, for ideal gases, equivalent to mole fractions. It is useful to keep in mind that volume fractions are not directly equivalent to mass fractions, especially when a species' molar mass differs significantly from the mean molecular mass of the atmosphere. The confusion can introduce significant differences in reported individual species' "partial" pressures.

Some publications track atmosphere compositions as equivalent pressures, that is the pressure a given species would have if it was the only component in the atmosphere. In that case, it is effectively a proxy for mass. Atmosphere evolution studies and models (escape mechanisms, for example) often use this notation when the evolution of a few species is followed. It should not be confused with partial pressures that, strictly speaking, require that total pressure and each individual species would be tracked. The distinction is sometimes not stated explicitly. 

With that in mind, water abundances are also referred to as fractions of a terrestrial ocean (TO) or as Global Equivalent Layer (GEL) depth - the equivalent depth of the condensed water layer, neglecting the planet's topography. In such a case, one terrestrial ocean (1.4~$\times$~10$^{21}$~kg) as water vapour corresponds to an equivalent surface pressure of 268 bar and a GEL of about 3000 m, while the partial pressure still depends on the composition of the rest of the atmosphere. At present-day, water in the atmosphere of Venus corresponds to an equivalent pressure of about 2~mbar or a $\approx$ 2 cm GEL.

\subsection{Isotopes and noble gases}

The elemental and isotopic compositions of volatile elements (C, H, O, N, S and noble gases) contained in the Venus atmosphere hold clues on the entire geological history of Earth's sister planet and have already been reviewed elsewhere \citep[\textit{e.g.,}][]{von_zahn_composition_1983,donahue_origin_1983,wieler_noble_2002,johnson_venus_2019,chassefiere_evolution_2012,avice_perspectives_2020,avice_noble_2022,zahnle2023}. Only some significant facts and research avenues in light of recent discoveries for Earth and Mars are highlighted and discussed below.

\subsubsection{The deuterium-hydrogen ratio}
The D/H ratio (where D$=^{2}$H) of hydrogen in the Venus atmosphere, with a $\delta$D$_{VSMOW}$ value close to 120,000\textperthousand, is much higher than all other known reservoirs in the Solar System. This extreme enrichment in deuterium has been interpreted as evidence that Venus suffered from hydrogen escape during its geological history \citep{donahue_venus_1982}. While there is little doubt that hydrogen escape had a strong influence on the entire geological history of Venus \citep{baines_atmospheres_2013}, the fact that both the starting D/H ratio and the regime of escape remain unknown prevent us from reconstructing a precise history of hydrogen (and water) escape from Venus. 

Recently, \citet[][]{zahnle2023} proposed a new model of diffusively-limited hydrodynamic escape in the context of loss of water from an early steam-reach atmosphere for early Venus. For this type of loss mechanism, escape of hydrogen originally located in water molecules would lead to changes in the D/H ratio, but also in the mass-dependent and mass-independent isotopic composition of atmospheric oxygen. Importantly, all model results point toward an enrichment of the D/H ratio on the order of 100, which confirms that $\delta$D measurements alone are not sufficient to determine the escape history of planetary atmospheres. If hydrodynamic escape was vigorous enough, neon and argon could also suffer from atmospheric escape and be lost from the atmosphere \citep[e.g.][]{Gillmann2009}. The most important outcome of this model is that new precise measurements of the Ar/Kr ratio in the Venus atmosphere \citep[][]{avice_noble_2022} could be decisive to determine whether Venus suffered from extreme escape (loss of Ar and fractionation of the Ar/Kr ratio) or from minimal escape.

\subsubsection{Neon and Argon}
\begin{figure*}
\centering
  \includegraphics[width=0.75\textwidth]{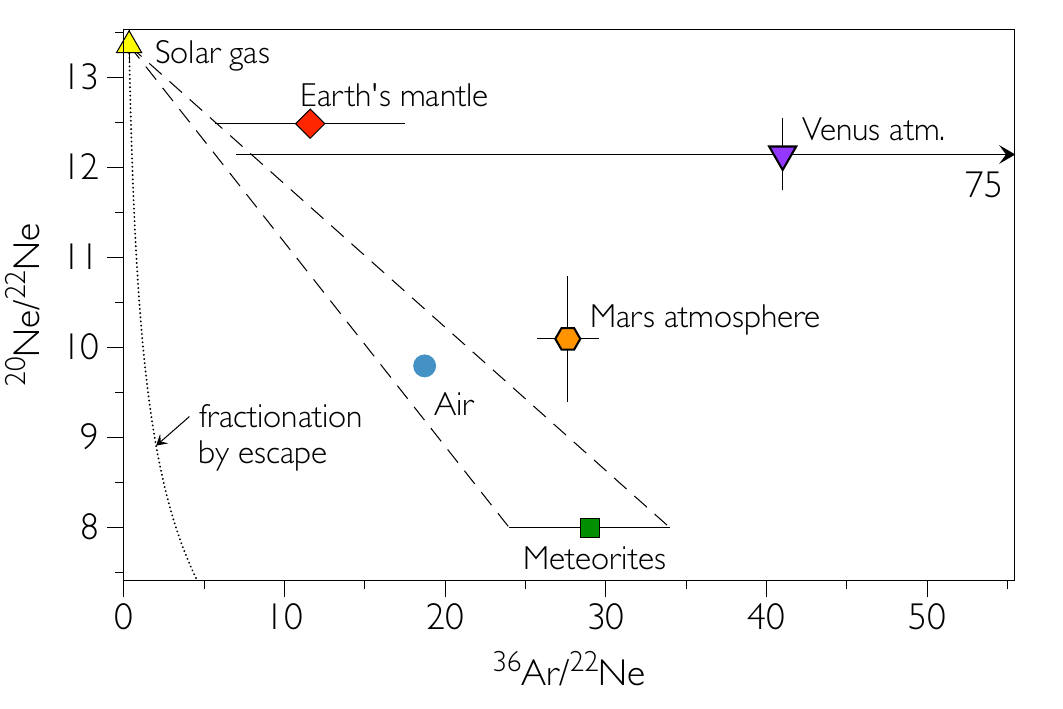}
\caption{Ne-Ar isotope diagram. The isotopic composition of Earth atmospheric neon lies on a mixing range between Solar and meteoritic end-members. The current estimate for Venus atmospheric neon suggests that, similarly to Mars atmospheric neon, Venus could lie outside this mixing range, although the uncertainty on the Ar/Ne ratio in the Venus atmosphere is high. Figure modified after \citet{marty_origins_2012,avice_noble_2022}.}
\label{fig:NeonIsotopes}       
\end{figure*}

For noble gases, measurements of neon and argon by previous space missions allowed to propose several interesting working hypotheses for the origin and evolution of the Venus atmosphere (Fig. \ref{fig:NeonIsotopes}). First, neon and argon are about one order of magnitude more abundant in the atmosphere of Venus compared to the Earth's atmosphere. Secondly, the Ne/Ar ratio is close to chondritic proportions (see \citet{avice_noble_2022} and refs. therein). While this could suggest that chondritic bodies delivered more volatile elements to Venus compared to Earth, the $^{20}$Ne/$^{22}$Ne ratio is close to 12, a value much higher than measured for neon trapped in chondritic meteorites and much closer to the value measured for solar neon. Several scenarios could be envisaged in order to explain this conundrum. 

A first scenario would consist in simply mixing chondritic with solar neon (and argon) to explain the elemental and isotopic composition of these two elements in the atmosphere of Venus. This would lead to an isotopic ratio of neon intermediate between the two end-members aforementioned, similarly to the case of Earth \citep{marty_origins_2012, zhang_noble_2023}. While the Ne/Ar ratio would also deviate from a starting solar-like value, its current determination is not precise enough to evaluate whether Venus plots off the mixing trend between the two end-members (see Fig. \ref{fig:NeonIsotopes}). A shift in the $^{38}$Ar/$^{36}$Ar ratio due to the addition of chondritic argon to solar argon would be very difficult to measure, given the fact that solar and chondritic argon have very similar $^{38}$Ar/$^{36}$Ar ratios.

An alternative scenario proposed by \citet{avice_noble_2022} would involve a delivery of volatile elements by cometary bodies in addition to the two-component mixing invoked above. Comets are rich in argon and probably devoid of neon so that such a cometary contribution would only impact the Ar/Ne ratio. New precise measurements of the elemental ratio Ar/Ne in the Venus atmosphere and of the isotopic composition of xenon (see below) would definitively shed new light on the origin of noble gases (and all volatile elements) in the Venus atmosphere.

\subsubsection{Xenon}
\begin{figure*}
\centering
  \includegraphics[width=0.75\textwidth]{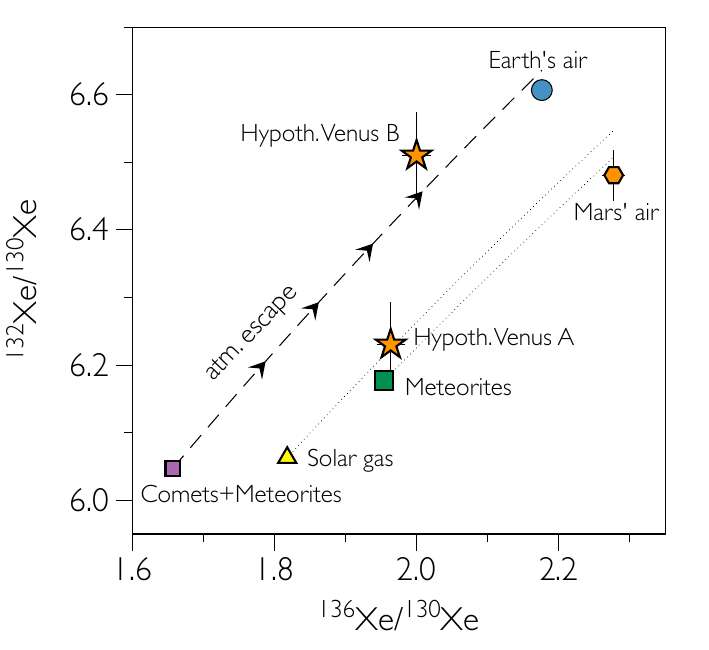}
\caption{Three isotope diagram of xenon isotopes. Earth's atmospheric xenon is enriched in heavy isotopes and plots on a mass-dependent fractionation curve starting at U-Xe (a mixture between cometary and meteoritic gases). Mars' atmospheric xenon is also enriched in heavy isotopes, but its starting composition is closer to the Solar and chondritic end-members. Two completely hypothetical isotopic compositions for Venus atmospheric xenon are also plotted (Hypoth. Venus A and B). For hypothetical Venus A, xenon would be less fractionated than on Earth and Mars and it would be challenging to decipher what is the starting isotopic composition (Chondritic or Solar). For hypothetical Venus B, a cometary contribution similar or slightly higher than for Earth would explain the low $^{136}$Xe/$^{130}$Xe ratio and the extent of isotopic fractionation would be lower than for terrestrial atmospheric xenon. These two examples highlight the power of xenon isotope systematics to reveal information about the origin and evolution of the Venus atmosphere. Of course, many other hypothetical compositions for Venus' atmospheric xenon could be imagined.}
\label{fig:XenonIsotopes}       
\end{figure*}

The isotopic composition of xenon contained today in the atmosphere of Venus is currently unknown but has the potential to unveil major information regarding the origin and evolution of the Venus atmosphere \citep[see details in][]{avice_noble_2022}. A comparison with Earth's and Mars' xenon is relevant here to understand the potential of such measurements.
On Earth, the starting composition of atmospheric xenon is selectively depleted in its two heaviest isotopes, $^{134}$Xe and $^{136}$Xe compared to solar and chondritic end-members. This feature led to the definition of U-Xe, the progenitor of atmospheric xenon \citep{pepin_origin_1991}. Measurements of xenon emitted by the comet 69P/Churyumov-Gerasimenko revealed that cometary xenon is even more strongly depleted in $^{134}$Xe and $^{136}$Xe compared to U-Xe \citep{marty_xenon_2017}. U-Xe would thus consist of a mixture of chondritic (78\%) and cometary (22\%) xenon. The high abundance of noble gases in cometary gases compared to water \citep{balsiger_detection_2015} implies that comets could have delivered a significant budget of atmospheric noble gases without altering the chondritic-like D/H ratio of terrestrial water \citep{bekaert_origin_2020}. The case of Mars is different as analyses of gases contained in martian meteorites and in-situ measurements by the Curiosity rover point toward a pure solar origin for Mars atmospheric xenon \citep[][and refs. therein]{conrad_situ_2016,avice_noble_2018}. Precise measurements of xenon isotopes in the Venus atmosphere (see Fig. \ref{fig:XenonIsotopes}) could thus help better understand what was the delivery mix (solar/chondritic/cometary) of volatile elements to Earth's sister planet.

Atmospheric xenon on Earth and Mars also presents two striking features. Its abundance is low compared to what would be expected if it was sourced by solar gas, chondritic bodies or comets, and it is enriched in heavy isotopes and depleted in light isotopes relative to all known cosmochemical starting end-members. In other words, xenon is depleted in abundance by about one order of magnitude and isotopically fractionated in favor of heavy isotopes. Original studies proposed mechanisms such as hydrodynamics escape on early Earth ($<$100~Myr) to selectively deplete (and fractionate) xenon from the Earth's atmosphere \citep[reviewed by][]{dauphas_geochemical_2014} until the discovery that the isotopic composition of atmospheric xenon has been evolving until about 2.5~Ga ago, at the time of the Great Oxidation Event \citep[see][for a recent curve for the evolution of atmospheric xenon]{ardoin_end_2022}. This isotopic evolution, originally discovered by \citet{pujol_chondritic-like_2011}, could be related to prolonged loss of xenon from the Earth's atmosphere to space accompanied by the progressive enrichment in heavy isotopes of the remaining fraction \citep{avice_evolution_2018}. \citet{zahnle_strange_2019} proposed a model in which a flux of partially ionized hydrogen would escape from the Earth's atmosphere and, via Coulomb interactions, would drag xenon ions. This mechanism of escape requires having a total mixing ratio of hydrogen exceeding 1\% and special and still underconstrained conditions to bring xenon ions to high altitudes, where hydrogen escape happens. While there are still many uncertainties regarding this escape mechanism, the evolution of the isotopic composition of terrestrial atmospheric xenon could be a tracer of at least a portion of the history of hydrogen escape from planetary atmospheres \citep{avice_perspectives_2020}. 

Isotopic fractionation during escape should be accompanied by a progressive depletion in abundance of xenon compared to other noble gases. \citet{broadley_high_2022} recently reported a high Xe/Ar ratio for gases extracted from $>$3~Ga-old hydrothermal quartz crystals from the Barberton Greenstone Belt (South Africa) and interpreted such measurements as evidence that xenon was indeed still escaping from the Earth's atmosphere during the Archean. As mentioned above, Mars atmospheric xenon is also isotopically fractionated and the fractionation is of the same order of magnitude as for Earth's atmospheric xenon. However, the timing for the isotopic fractionation of xenon on Mars appears to be different. Measurements of ancient Martian atmospheric xenon trapped in Martian meteorites suggest that the isotopic composition of atmospheric xenon measured today was already established circa 4~Ga ago \citep{cassata_xenon_2022}. Under favorable conditions, the joint atmospheric escape of hydrogen and xenon on early Mars could be viable, these conditions ceased to operate some time during the Noachian. Planned measurements of the abundance and isotopic composition of atmospheric xenon on Venus will certainly help us to understand the role of hydrogen escape in shaping the atmosphere of Venus we see today \citep{garvin_revealing_2022}.

\subsubsection{Krypton}
The case of krypton is also important and often overlooked. The abundance of krypton in the atmosphere of Venus remains poorly constrained (see the recent review by \citet{{avice_noble_2022}}) and its isotopic composition is unknown, but recent studies on Earth and Mars krypton suggest that krypton is also a powerful tool to understand the origin of planetary atmospheres. \citet{peron_deep-mantle_2021} reported measurements of deep mantle krypton sampled in the volcanic rocks produced by the Galápagos and Iceland plumes. Mantle krypton seems to have been sourced by chondritic bodies but \citet{peron_deep-mantle_2021} also reported a negative anomaly in neutron-rich $^{86}$Kr suggesting that none of the meteorites present in our collections are representative of the chondritic bodies which delivered volatile elements to the Earth's interior. For Mars, \citet{peron_krypton_2022} reported measurements of krypton contained in the Chassigny meteorite. Results suggest that chondritic krypton is also present in Mars' interior while krypton in the martian atmosphere has been sourced by solar gases \citep{conrad_situ_2016}. Future measurements of the isotopic composition of krypton in the atmosphere of Venus thus have the potential to improve our understanding of the nature of the bodies which brought volatile elements to terrestrial planets.

\subsubsection{Radiogenic isotopes}
\label{sub:arg}
Excesses of radiogenic and fissiogenic isotopes of noble gases originally produced in the silicate parts of Venus and later degassed into its atmosphere can potentially inform us about the intensity and timing of degassing, and thus on the geodynamics of Venus. Venus' $^{40}$Ar measurements have been used to estimate that only 10 to 34\% of the total $^{40}$Ar inventory produced over 4.5~Gyr has been outgassed into the atmosphere \citep{kaula_constraints_1999,orourke_thermal_2015}, while the estimate reaches about 50\% for the case of terrestrial argon \citep{allegre_rare_1987}. Venus outgassing could have been limited during most of its evolution, or was only important during its early history, when $^{40}$Ar had not formed yet. However, these estimates depend highly on estimates of the K/U ratio for bulk silicate Venus \citep{orourke_thermal_2015}. Current constraints come from surface measurements of the abundances of K and U by the Vega and Venera missions \citep{Surkov1987}. In the argon budget mentioned above, the K/U ratio was estimated at $\approx$ 7,000 \citep{orourke_thermal_2015}, but an Earth-like value closer to 16,000 \citep{arevalo_ku_2009,zhang_noble_2023} would reduce the inferred portion of Ar to $\approx$ 10\% \citep[see also][]{Gillmann2022}. Future measurements of the fissiogenic excesses of xenon isotopes produced by both extinct ($^{129}$Xe, $^{244}$Pu) and extant ($^{238}$U) radionuclides and present in the atmosphere would help combining radiogenic isotope systematics (K-Ar, Pu-U-Xe) and ultimately contribute to reconstruct more precisely the degassing history of Venus.

\subsection{Greenhouse effect and structure}

The thick atmosphere of Venus causes a greenhouse effect that raises its surface temperature to an average of approximately 735 K, with spatial variations of up to 100~K, depending on the topography. While Venus does receive about twice the Earth's sunlight, with a solar constant of 2622~W~m$^{-2}$ vs 1364~W~m$^{-2}$, the planetary (Bond) albedo of Venus is much higher than Earth's (0.3) and is typically taken as 0.76, although values of 0.7-0.82 \citep[][]{colin1983} and up to 0.9 \citep[][]{mallama2006} have been suggested, and radiative-convective equilibrium models are consistent with an albedo of 0.81 \citep[][]{bullock2017}. The role of clouds is particularly important in the thermal and radiative structure of the atmosphere of Venus \citep[a complete review can be found in][]{Titov2018}. About three-fourths of the incoming sunlight is reflected by Venus' clouds back to space. Only about 11\% of the total absorbed sunlight reaches the surface of Venus, as it is absorbed on the way by CO$_2$ and an unknown UV absorber \citep[][]{titov2007,titov2012,limaye2018venus}. This absorber is presumably sulfur-based, given the composition of the atmosphere of Venus, and has been suggested to be droplets or particles in the clouds or sulphur-based polymers \citep[][and references therein]{pollack1979,Pere18,frances2022}, the main candidates being S$_2$O$_2$ and S$_2$O. The absorber has also been suggested to be derived from iron-sulfur chemical reactions \citep{jiang2024}, even though Venus' atmospheric chemistry remains poorly understood \citep[e.g.][]{Bier20}. The precise vertical distribution of the absorber around 70~km altitude, both above or below the cloud top \citep[i.e.][]{lee2021} also remains debated. Work on this topic involves the chemistry of sulfur species and the reactions between polysulfur molecules, but is hindered by the lack of comprehensive lab experiments to determine equilibrium constants for these equilibria, given how dangerous and expensive they can be.

Similarly, very little black body radiation from the surface can escape due to the high IR opacity of the main greenhouse species in the lower atmosphere (CO$_2$, H$_2$O, SO$_2$). Instead, the planet radiates from the colder layers at a higher altitude and its equilibrium temperature is 232~K, lower but comparable to Earth's 254 K. CO$_2$ has been suggested to be responsible for a temperature elevation by about 300~K due to the greenhouse effect \citep[][]{pollack1980}, while the rest of the greenhouse effect is attributed to SO$_2$ and water despite their low relative abundances. 

The thermal structure of the atmosphere is a consequence of its composition, absorption and emission of radiation by the different species, clouds and convection processes. As a result, temperature varies with the altitude \citep[see][for a full review and observational history]{limaye2018venus}. In the lower atmosphere, defined as the troposphere, below 60~km, temperature decreases with altitude, from 735~K at the surface to 245~K at the tropopause, nearly following the (dry) adiabatic lapse rate at about -8~K.km$^{-1}$.
Venus lacks a stratosphere due to the low maximum concentration of ozone in the atmosphere (see Section \ref{sub:chem}), which is insufficient to form an Earth-like ozone layer that would heat up the corresponding layer and induce a reversal of the temperature gradient with altitude. Instead, above the clouds (70-150~km) the mesosphere has a more complex structure with a succession of layers of alternating warm and cold temperatures \citep[][]{limaye2018venus}.

The thick atmosphere also imposes a uniform temperature between day and night, at the surface, at constant altitude, while in the upper atmosphere, above $\approx$ 100~km, day/night temperature variations are of the order of 100 K. Variations of temperature with longitude and latitude are relatively small, especially in the lower atmosphere, below $\approx$ 60~km.

\subsection{Dynamics}

The dynamics of the atmosphere of Venus are distinct from Earth's in several ways due to planetary properties. (1) The rotation of the solid body of Venus is slower by more than two orders of magnitude than the Earth's. The atmosphere is then in cyclostrophic balance, meaning that the Coriolis force is negligible and that the pressure gradient force is balanced by the centrifugal force. (2) The axial tilt of the Venus is less than 3$^{\circ}$, leading to no seasonal variability. (3) The absence of a stratosphere and (4) the complete cloud coverage are two other features that will induce more complex differences in atmospheric dynamics compared to Earth. 

\subsubsection{Large-scale circulation and super-rotation}
Venus' atmosphere is in super-rotation, a specificity it shares with Saturn's largest moon Titan. It is rotating globally faster, approximately 60 times at cloud top, than the solid body of the planet \citep[][]{Boye69}. The atmospheric angular momentum budget and redistribution need to be quantified to understand the atmospheric circulation. In a nutshell, the super-rotation depends on the  counterbalance between the transport of angular momentum by the mean meridional circulation and by transient wave activity \citep[][]{Lebo20}. On Venus, the Hadley-cell-like mean meridional circulation, through an equator-to-pole thermally direct circulation, is transporting angular momentum upward and poleward. This transport tends to homogenize the distribution of the angular momentum. In order to maintain the super-rotation of the atmosphere, this momentum transport needs to be counterbalanced by one or several others transport mechanisms.

Different hypotheses regarding the nature of such a mechanism are in question, such as vertical transport by eddies (like thermal tides) and horizontal equator-ward transport by planetary-scale waves \citep[][]{Gier75,Ross79}. Numerical simulations were performed using different atmospheric general circulation models, and super-rotation was obtained in these models \citep[][]{Lebo13,Sugi19}. In those simulations, the angular momentum transport is balanced by tidal \citep[][]{Lebo16} and other waves generated by hydrodynamic instabilities, such as the barotropic \citep[][]{Lee10,Lebo16,Mend16}, baroclinic \citep[][]{Lebo16,Sugi19}, and Rossby-Kelvin instabilities \citep[][]{Mend16}, as well as turbulent eddies \citep[][]{Sugi19}. 

\begin{figure*}
\centering
  \includegraphics[width=1.\textwidth]{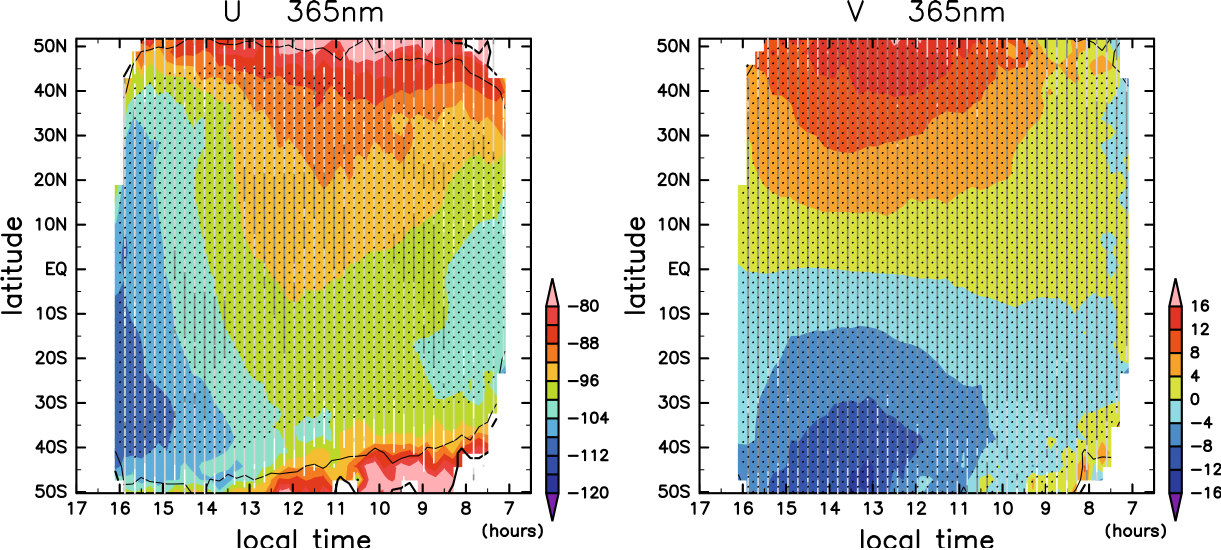}
\caption{Mean zonal winds (left) and meridional (right) as functions of local time and latitude for the period December 2015–March 2017 measured by Akatsuki's Ultraviolet Imager (UVI) at 365~nm \citep[][]{Hori18}. Adapted from \cite[][]{Hori18}.}
\label{fig:Dyn}       
\end{figure*}

Using cloud tracking at the cloud top region from the Akatsuki spacecraft's Ultraviolet Imager (UVI), the horizontal and vertical angular momentum transport by thermal tides, planetary-scale Rossby waves, and other transient disturbances were estimated. The thermal tides, atmospheric pressure variation due to the diurnal differential heating of the atmosphere, appear to transport angular momentum towards the equator, contributing to the maintenance of the super-rotation against the homogenization by the meridional circulation \citep[][]{Hori20}. Maps at cloud-top altitudes of the zonal and meridional averaged over several Venus days \citep[][]{Hori18}, showed local time dependence consistent with known tidal features (Fig~\ref{fig:Dyn}). With the same methodology but in the lower and middle cloud layers using Akatsuki/IR2 camera, an equatorial atmospheric jet current was observed \citep[][]{Hori17}. Though the mechanism to generate the equatorial jet is not yet fully understood, downward momentum transport from thermal tides is one of the main hypotheses.

\subsubsection{Interaction with the surface}
The interaction between the surface and the atmosphere is a major aspect of exchanges of heat and angular momentum, which impacts the thermal and wind shear and, thus, the atmospheric dynamics and the rotation of the solid body itself. On Venus, due to the technological difficulty in probing this region below the cloud layer and landing spacecraft, the first 10~km region above the surface remains largely uncharted. Only a limited amount of data have been collected with a few probes. The VeGa-2 probe has measured the only temperature profile in that region \cite[][]{Link86b}. In this temperature profile, the atmosphere appears to be highly unstable at altitudes below 7~km, which is highly unrealistic under the usual assumptions. This unexpected temperature profile could be explained by a change in the atmospheric molecular mass and gas composition in this region. The near-surface atmosphere is assumed to be a homogeneous layer of CO$_2$ and N$_2$, however with a gradual decrease of N$_2$, from 3.5\% to near-zero at the surface due to supercritical fluid properties, the measured temperature profile would be well-mixed and therefore plausible \cite[][]{Lebo17,Lebo20b}. 

At the surface, only Venera 9 and 10 directly measured the wind speed for less than an hour \cite[][]{Avdu77}, and several indirect measurements were performed by Venera 13 and 14 \cite[][]{Ksan83}. The amplitudes of the measured wind speeds are less than 2~m~s$^{-1}$ below 100~m \cite[][]{Lore16}, with values below 0.5~m~s$^{-1}$ as the most probable value. The height and dynamics of the planetary boundary layer (PBL) was not measured. Only two dunes fields have been observed in radar measurements with Magellan \cite[][]{Gree92}. 

Global Circulation Models simulations showed a correlation between the diurnal cycle of the planetary boundary layer activity and the topography and the diurnal cycle of surface winds \cite[][]{Lebo18}. The convergence of upward anabatic winds induces upward vertical winds in a deeper PBL depth at noon. The dynamics of the PBL was studied with a turbulent-resolving model \cite[][]{Lefe22c}, with simulation considering noon and midnight situations in both the equatorial plain and on Aphrodite Terra. A strong diurnal cycle in the high terrain was found, with a convective layer reaching up to 7~km above the local surface, with vertical speeds of 1.3~m~s$^{-1}$ at maximum. The boundary layer depth in the low plain is consistent with the observed wavelength of the dune fields. At noon, the amplitude of the surface wind field for both locations is strong enough to lift dust particles and engender micro-dunes. In the high terrain, the convergence of strong convection cells generates convective vortices, resolved in simulations for the first time.

The exchange between the surface and the cloud region is another puzzle of the Venusian atmosphere. Akatasuki Long-wave InfraRed camera (LIR) measured a large stationary bow-shape wave above the main equatorial topographic features at cloud top \citep[][]{Fuku17,Kouy17}. The largest waves extend over 60$^{\circ}$ of latitude and lasted over 10 Earth days. The majority of those features were observed in the late afternoon. They were then studied using numerical models. The main hypothesis is that mountain waves would propagate up to cloud-top altitudes. \citet[][]{Lefe20} developed a model able to resolve those type of waves. Over the largest equatorial mountains, waves with similar amplitude and latitudinal extent were present. Generated by the mountain forcing the flow to pass over them, waves are propagating upward, favored by the near-surface atmosphere more stable in the afternoon. The waves encounter two neutral layers, the deep-atmosphere mixed layer between roughly 18 and 30~km, and the cloud convective region between around 50 and 60~km. In these two regions, additional trapped lee waves are generated, propagating horizontally over several hundreds of kilometers. The mountain waves are observed in the cloud top but would propagate above, where the zonal wind changes sign. Such waves were implemented in the IPSL (Institut Pierre Simon Laplace) Venus Planetary Climate Model (PCM) \citep[][]{Nava18}, and their effect on the angular momentum could cause a decrease in the length of day. Waves propagate into the thermosphere and dissipate via viscosity \citep[][]{Hick22}, with possible large effects on momentum and temperature. The impact of the mountain waves on chemistry as a possible vertical mixing mechanism is not known. 

\subsection{Chemistry and clouds}

\subsubsection{Chemical cycles}

The atmospheric chemistry on Venus is dominated by three main large-scale cycles: the CO$_2$ cycle, sulfur oxidation, and polysulfur cycles \citep[][]{Mill07,Marc18,Bier20}. The carbon dioxide cycle is composed of the photolysis of CO$_2$ on the day side producing CO and atomic oxygen, the transport to the night side of a part of these products, then the formation of O$_2$ observed with O$_2$ airglow, and to finish the catalytic conversion of CO and O$_2$ into CO$_2$. 

The sulfur oxidation cycle consists of the upward transport of OCS (Carbonyl sulfide) and SO$_2$, the formation of H$_2$SO$_4$ by the oxidation of OCS and SO$_2$, the condensation of H$_2$SO$_4$ and H$_2$O to form the global cloud layers that will sediment and evaporate, and the thermal decomposition of H$_2$SO$_4$ to produce OCS and SO$_2$ \citep[][]{Zhan12,Bier20,Shao20}. 

The polysulfur cycle involves the upward transport of OCS and SO$_2$, the production of molecular sulfur by photodissociation and the production of polysulfur (Sx) via catalytic processes, the downward transport of these products and their thermal decomposition, and then the production of OCS from reaction with CO and of SO$_2$ by oxidation \citep[][]{Zhan12,Kras13}. There are strong measurements that support the carbon dioxide and sulfur oxidation cycles, whereas the polysulfur cycle remains speculative.

\subsubsection{The cloud layer vertical structure}

Venus is completely enshrouded by a cloud layer between 45 and 70~km of altitude. The Venera, Pioneer Venus (PV), and Vega descent probes measured the composition and vertical structure of the clouds. The droplets at the equator are composed on average of 80\% of H$_2$SO$_4$ and of 2\% H$_2$O, both present in liquid form. The weight fraction of sulphuric acid in droplets increases up to 90\% at 60$^{\circ}$ and then decreases 75\% at the poles. Three main cloud particle size modes were measured by PV-LCPS (Large Probe Cloud Particle Size Spectrometer) with different vertical distribution: 
\begin{itemize}
    \item Mode 1 is composed of spherical particles with a mean radius of 0.30~$\upmu$m, it is the most abundant mode in the clouds, present throughout the clouds, but more abundant in the lower and upper part of the clouds.
    \item Mode 2 is composed of particles with a mean radius of 1.0~$\upmu$m, present throughout the clouds, and more or less constant vertically
    \item Mode 3 is composed of non-spherical larger particles with a mean radius between 3.0 and 4.0~$\upmu$m, present only in the lower and middle cloud layers. The existence of this mode is controversial, and could be the “tail” of mode 2 instead of a separate mode
\end{itemize}

To fit the measured IR opacities, some models add a mode 0 and a mode 2', with particles' mean radius of 0.06 and 1.4~$\upmu$m, respectively. 
From the vertical distribution of these different modes, the cloud layer is considered to be vertically stratified in 3 regions: the lower cloud from 47.5 to 50.5~km, the middle cloud from 50.5 to 56.5~km, and the upper cloud from 56.5 to 70~km. The cloud layer is bordered by haze layers: the upper haze layer ranging from 70 to 90~km with particles close to mode 1, and the lower haze layer lower than 30~km altitude for which few characteristics are known. The bottom floor of the cloud layer appears almost constant for all latitudes, whereas the cloud top varies from 72~km at the equator to 62~km near the poles. The PV-LCPS observations gave an estimate of the vertical distribution of droplets, however little is known about the local time and latitudinal variability. Akatsuki also observed mesoscale transient features in the lower and middle cloud layers \citep[][]{Pera19,Pera20,Hori23}, suggesting active dynamics at several scales that need to be studied.

\subsubsection{The cloud layer chemistry}
\label{sub:chem}

\begin{figure*}
\centering
  \includegraphics[width=1.\textwidth]{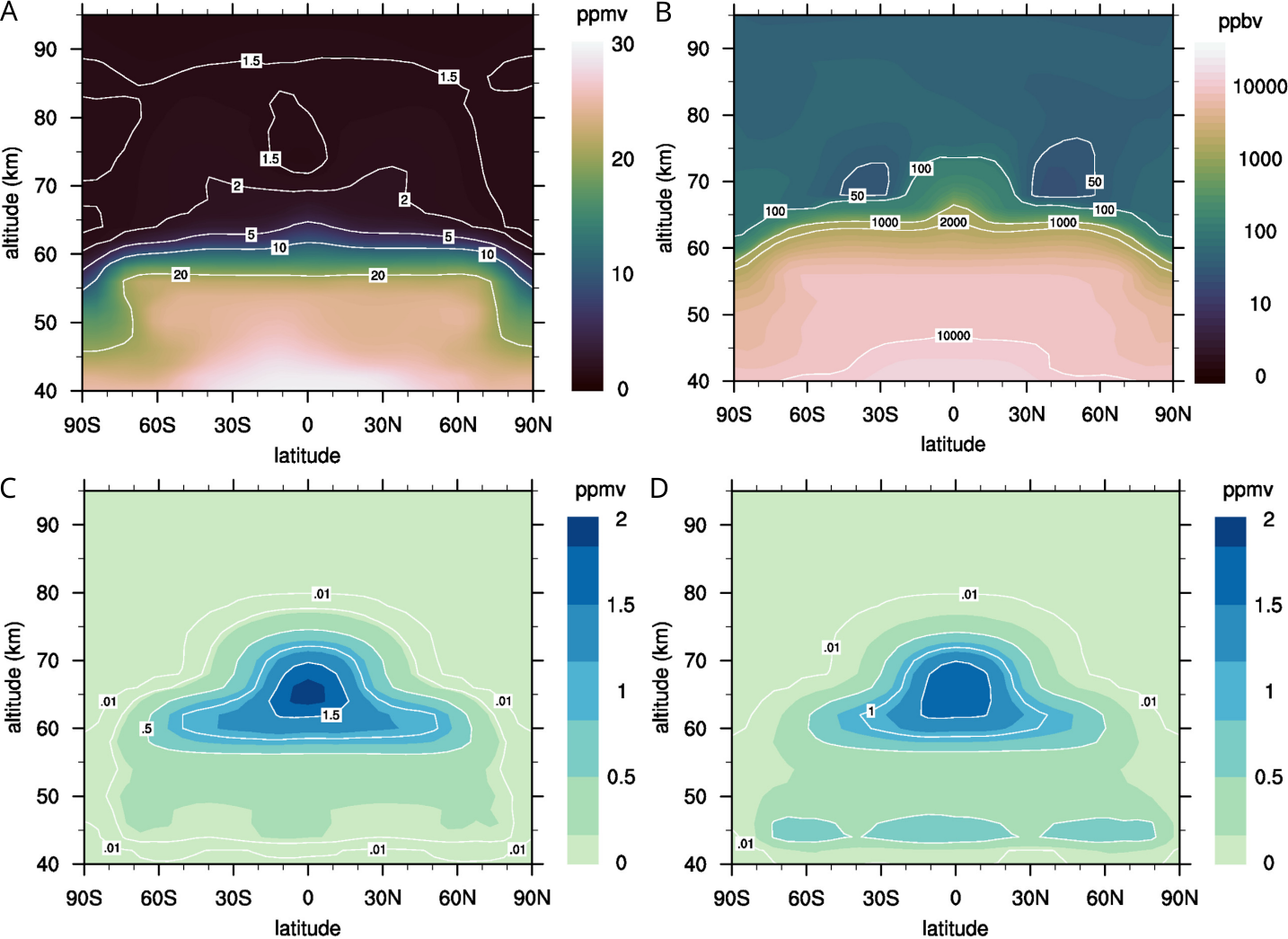}
\caption{Zonally-averaged distribution of the gas-phase (A) and liquid phase (C) of H$_2$O and the gas-phase (B) and liquid phase (D) of H$_2$SO$_4$ obtained with the IPSL Venus PCM \citep[][]{Stol23}. The outputs of the model are available online through the Venus Climate Database (VCD), see \url{http://www-venus.lmd.jussieu.fr}, where a climatology with mean values and spatial and temporal variability of many dynamics and chemical fields of the Venusian atmosphere from the surface to the exosphere, validated against available observations, is provided from the latest version on the model \cite[][]{Martinez2023}. Adapted from \cite[][]{Stol23}.}
\label{fig:Chem}       
\end{figure*}

The recent long-term mission Venus Express was able to measure the chemistry in the clouds layer and above for almost a decade. With SPICAV-UV/Venus Express nadir \citep[][]{Marc20}, cloud-top SO$_2$ was measured to be decreasing with increasing latitude, from typically 5–100~ppbv at 70~km near the equator to 2–20~ppbv polewards of 60$^{\circ}$N and S. The IPSL Venus PCM enlightens about the interaction between the dynamics and chemistry \citep[][]{Stol23}, with the first three-dimensional transport modeling of minor species in the atmosphere of Venus (Fig~\ref{fig:Chem}). From Venus Express data, the unknown UV absorber is comparatively brighter at 365~nm, i.e. less absorbent meaning less absorbing materials, at mid-to-high latitudes compared to lower latitudes \citep[][]{Titov2018} along with an already known darkening at low latitudes in the 12:00–14:00 local time range \citep[][]{titov2012}. A long-term UV darkening, i.e. increased absorption by the unknown absorber, was observed between the years 2006 and 2011. It was followed by a variable increase in UV darkness between 2011 and 2015. A similar trend was also measured by the Venus Monitoring Camera (VMC) instrument on board Venus Express \citep[][]{Lee19}.

At the equator, the cloud-top SO$_2$ abundance was observed to decrease throughout the mission duration, from 80~ppbv (in median) in 2007 to about 10~ppbv in 2014. This decrease was not monotonous, as several peaks were observed in 2009, 2012, and late 2013. Potential topographic correlation signatures of Aphrodite Terra were observed both in SO$_2$ and in UV brightness, with a high cloud-top SO$_2$ abundance and brighter cloud top above Aphrodite Terra, in the 30$^{\circ}$–45$^{\circ}$E zonal region.

The reanalysis of the SPICAV-UV nadir dataset through UV absorption near 250~nm in the backscattered solar light highlighted a permanent ozone layer on Venus near 70~km above latitudes of 50$^{\circ}$ \citep[][]{Marc19}. The ozone layer originates from the downward transport of O$_2$ (about 50~ppmv) over the poles by the mean meridional circulation \citep[][]{Stol23}. In the Northern Hemisphere, O$_3$ mixing ratio is maximal around 80$^{\circ}$N, in the strong temperature inversion named cold collar, with mean values of 20~ppbv, i.e. about a thousand times less than on Earth. In the Southern Hemisphere, the peak is around 60$^{\circ}$S, with values around 10~ppbv. A local time variability was also observed, with more O$_3$ near the morning terminator than near the evening terminator.

For a decade, SO$_2$ and HDO (proxy for H$_2$O) at cloud-top altitude has been monitored with the TEXES (Texas Echelon Cross Echelle Spectrograph) at NASA's InfraRed Telescope Facility (IRTF, Mauna Kea Observatory) where long-term and short-term observations were performed. An anti-correlation is visible in the long-term variations of H$_2$O and SO$_2$ at the cloud top altitude. Since the year 2015, a long-term decrease of H$_2$O, associated with a long-term increase of SO$_2$ \citep[][]{Encr20b}, was observed. SO$_2$ mixing ratio shows also strong short-term temporal and spatial variations, suggesting the occurrence of SO$_2$ plumes with a lifetime of a few hours. On the other hand, the H$_2$O abundance is almost spatially uniform and shows only few variations in function of time \citep[][]{Encr19}. The SO$_2$ plumes are mostly occurring in the $\pm$~30$^{\circ}$ latitude region, and are less frequent at noon, compatible with local time variation and timescale from the cloud convective layer \citep[][]{Imam14,Lefe18, Lefe22a}. However, the origin of these plumes is not clear.

\subsubsection{Tropospheric chemical measurements}

The troposphere and lower clouds are difficult areas to probe, only a handful of study were performed below the clouds. \citet[][]{Marc23} measured a latitudinal increase of CO of about 30\% between the equator and 60$^{\circ}$N, with an anti-correlation with OCS. Ground measurements are then complementary tools, allowing for a follow-up of long-term trends. There is a particular lack of knowledge about the temporal and spatial variability of chemical species in the troposphere. SO$_2$ was observed between 30 and 40~km altitude using the NASA Infrared Telescope Facility (IRTF) and the iSHELL echelle spectrograph \citep[][]{Marc21}. A wide range of value was observed for the first time, with values between 120 and 220~ppm and strong latitudinal variation. The minimum is located at 15$^{\circ}$S, and it increases up to 40$^{\circ}$N, the farthest area observed. An increase between 30 and 40~km in altitude of SO$_2$ was also measured, consistent with the interaction with Hadley-cell circulation \citep[][]{Tsan08}.

\subsubsection{Discussion around the detection of phosphine}

Using JCMT (James Clerk Maxwell Telescope), \citet[][]{Grea21} claimed to have measured about 20~ppbv of phosphine in the clouds of Venus, at least 6 orders of magnitude higher than the expected abundance. With the NASA/DLR Stratospheric Observatory for Infrared Astronomy (SOFIA), a subsequent observation of 1~ppbv of phosphine at 75~km was claimed \citep[][]{Grea22}. Phosphine is present on Earth, associated with anthropogenic activity or microbial presence, and can also be found in reducing atmospheres of giant planets. On Venus, phosphine could have several origins \citep[][]{Grea21}, like volcanism, lightning or microbial organisms. Following the detection claim, a new analysis of the in-situ archival data from the Pioneer Venus Large Probe Neutral Mass Spectrometer (PV-LNMS) was performed \citep[][]{Mogu21}, that seemed to support the presence of phosphine. However, several studies have challenged this measurement \citep[][]{Snel20, Thom21,Vill21, Akin21,Linc21}, suggesting the pollution by SO$_2$ or disputing the claimed altitude of the phosphine detection. The detection of phosphine by the Venus Express instrument SOIR showed an upper limit several orders of magnitude below the claimed abundance \citep[][]{Trom21}. With the TEXES imaging spectrometer, an upper limit at least four time lower \citep[][]{Encr20} was established. Additional ground observations and in-situ measurements are needed to clarify the presence of phosphine. Alongside with phosphine, ammonia potentially detected by Venera-8 \citep[][]{Mogu21} and PV-LNMS \citep[][]{Mogu21}, is another  biosignature candidate in the atmosphere of Venus. In association with hydroxide salts in the cloud droplets, ammonia has been proposed to efficiently dissolve and trap SO$_2$ in the cloud region \citep[][]{Rimm21,Bain21}, and therefore explain the sharp SO$_2$ decrease in the clouds that has not been reproduced by chemical models so far. Like for phosphine, additional measurements are needed to support such claims.

\section{Interior and surface}\label{sec:intern}

\subsection{interior structure and composition}
 The main available constraint regarding the interior of Venus is its mean density that is close to Earth's, but lower by 2\%, if the same bulk composition is assumed \citep[e.g.][]{ringwood1977,aitta2012,Dumoulin2017}.The internal structure of both planets has therefore been argued to be comparable. The slow rotation of Venus has long made it difficult to obtain the moment of inertia of the planet that could also be used to estimate its interior mass distribution. However, observations of the rotation of the planet covering the time span from 2006 until 2020 allowed the moment of inertia to be estimated at about 0.337~$\pm$~0.024. Consequently, a rough estimate of the core radius of 3500~$\pm$~500~km could be obtained \citep[][]{Margot2021}. Substantially similar values were obtained by \citet{Shah2022} and \citet{Dumoulin2017}, while, using Earth’s PREM model, \citet[][]{amorim2023} obtained a most likely core radius of 3000-3500~km. However, comparatively large variations in the results can derive from uncertainties in light elements abundances, such as sulfur, and viscosity. Current measurements of the tidal Love number k$_2$ cannot uniquely constrain the state and structure of the core \citep{xiao2021,Dumoulin2017}, but based on pressures reached at the center of the planet \citet[][]{amorim2023} judge it unlikely that Venus has a solid inner core. Based on these results, it can be expected that the planet features both an upper and lower mantle. However, due to the lower internal pressures inside Venus, important phase transitions like basalt to eclogite and ringwoodite to bridgmanite are shifted to larger depths than inside Earth \citep[][]{Rolf2018, Shah2022, Maia2023}. Based on the core size estimate, it is likely that no post-perovskite is present in the lowermost Venus mantle (\citet[][]{Cizkova2010}; \citet[][]{xiao2021};\citet[][]{Rolf2022}). 

Similarly, the internal composition of Venus remains so far mostly unconstrained due to the lack of meteoritic and return samples that could host mantle-derived xenoliths. With no samples or direct observations, the mean density of Venus is, again, the main constraint on the composition of its interior. The amount of volatile species in the planet's interior, and especially its mantle, is still debated. The interior of Venus has long been suggested to be relatively dry compared to Earth's because it seemed to lack an asthenosphere (see Section \ref{sub:surf}) and because its atmosphere is very dry \citep[][]{grinspoon1993,bullock2001}. However, as discussed above in Section \ref{sub:arg}, radiogenic Ar may indicate Venus has outgassed fewer volatiles than Earth. Additionally, recent work applying chemical equilibrium during outgassing (both during the early magma ocean phase and long-term volcanism) suggests that water, even if present, may not be able to outgas efficiently under a thick multi-bar atmosphere \citep[][]{gaillard2014,gaillard2022}. This is further discussed in Section \ref{sec:evol}.

\subsection{Magnetic measurements}
Orbiter observations indicate that any self-sustained magnetic field must be at least $<$10$^{5}$ weaker than Earth's present-day magnetic field \citep[][]{Phillips_Russell1987}, and, thus, is most likely absent. Given a similar structure and composition as the Earth, Venus would be likely to generate a comparable magnetic field at present-day, according to models, as long as vigorous convection is generated in the core, even with its slow rotation \citep[i.e.][]{orourke2018a}. Possible explanations for the lack of an intrinsic magnetic dynamo are that (i) the core is still at least partly liquid, but currently not convecting. This would be due to low heat flow across the core-mantle boundary, possibly due to a basal magma ocean preventing efficient core cooling \citep[][]{ORourke2020}, (ii) the core is stably stratified, and in the absence of an event leading to remixing such as a giant impact, is not convecting \citep[][]{Jacobson2017} and (iii) the core has already fully solidified \citep[][]{Dumoulin2017}. So far, no crustal remanent magnetization has been detected on Venus, however due to the planet's dense atmosphere all orbital measurements so far had to be performed from a larger distance than, for example, the magnetic field measurements in martian orbit that led to the discovery of a patchy remanent magnetic field on this planet (\citet[][]{Acuna1999}; \citet[][]{PlattnerSimons2015}). Thus, potentially remanently magnetized regions on Venus might have eluded detection so far \citep[][]{orourke2018a}.

\subsection{Inferences from surface observation}
\label{sub:surf}
Most of the data we possess that can be used to infer the state of Venus' interior comes from the observation of the planet's surface and crust, looking for expressions of interior dynamics and characteristics. Radar data obtained by the Venera 15, Venera 16 and Magellan missions show that only 8\% of the surface of Venus are covered by mountainous, strongly tectonically deformed highlands, called tesserae \citep[][]{bindschadler1991,ivanov2011}. However, contrary to Earth, a good correlation exists between gravity and topography data \citep[][]{Kiefer1986} which might indicate the absence of an asthenosphere (\citet[][]{Kiefer1986}; \citet[][]{Steinberger2010}; \citet[][]{Rolf2018}). In general, topography on Venus differs widely from Earth's, as it is unimodal compared to Earth's bimodal distribution that reflects the presence of both oceanic and continental crust. Based on this, it has been suggested that the highlands like Beta, Atla and Themis Regio are being actively supported by mantle plumes \citep[][]{SmrekarPhillips1991}. However, recent dynamic gravity and topography analyses suggest that the data could be consistent with a shallow low viscosity zone, analog to Earth's asthenosphere, with a thickness of a couple of hundred kilometers \citep[][]{Maia2023}. When drawing an analogue to Earth, the presence of such an asthenosphere on Venus could be the result of partial melting \citep[][]{Maia2023}. \citet[][]{saliby2023} also suggests that the interior of Venus is likely to exhibit a significant viscosity contrast between the upper mantle and the lower mantle. Viscosity profiles generally comparable to Earth are usually preferred, with small depth variation in the lower mantle \citep[][]{Cizkova2010,Rolf2018,Steinberger2010}. However, estimating absolute viscosity values remains difficult, and lower overall viscosities for a Venus' mantle that is warmer than Earth's have been proposed \citep[][]{McGregor2023}.

\subsubsection{Tesserae}
The composition, age, and formation history of these tesserae are still unknown. It is also uncertain whether all highlands constitute a single geological unit, or whether there are actually several different units among them. Tesserae have been proposed to be the oldest surface features on Venus (\citet[][]{IvanovBasilevsky1993}; \citet[][]{OgawaYanagisawa2014}; \citet[][]{IvanovHead2011}; \citet[][]{Kreslavsky2015}) and at least one of them has been suggested to have a felsic composition (\citet[][]{hashimoto2008,Gilmore2015}; \citet[][]{Gilmore2017,shellnutt2019}), in line with partial melting experiments using starting material similar to venusian basalts \citep[][]{wang2022}. In case this will be confirmed by further measurements, implications for the history of Venus would be considerable, since the presence of surface liquid water is usually considered as a requirement for the formation of large volumes of felsic rocks. However, one should note that other pathways can produce felsic rocks without liquid water in small quantities. In that case, estimating the volume of felsic material becomes critical. 

Additionally, recent work \citep[][]{nimmo2023} using viscous relaxation models argues that weak felsic material would be less likely to maintain the observed plateau-like topography of tesserae than mafic material, unless the surface heat flux on Venus is exceptionally low. Tesserae have also been proposed as analogues of ancient continent-like surfaces. However, numerical models suggest that they do not behave as Earth's continental plates \cite[][]{Karlsson2020}. With  Alpha Regio as target region of the upcoming mission DAVINCI, we can expect valuable new data on tessera surface composition.

\subsubsection{Volcanic plains and the age of the surface}
The largest part of the Venus surface is covered by rolling lowland plains. Surface composition measurements performed by the Soviet Venera and Vega landers indicate that these plains consist of basaltic rocks similar to those formed at Earth's mid-ocean ridges \citep[see][for details on surface composition measurements]{fegley2014venus}, however, composition uncertainties remain large. The morphology of the surface in the plains is consistent with basaltic compositions \citep{Head1992}.

Based on the sparse cratering record of the planet, the surface is suggested to be young, ranging in age from 250 to 1000 Ma (e.g. \citet[][]{McKinnon1997}; \citet[][]{Schaber1992}; \citet[][]{Strom1994}; \citet[][]{le2011}; \citet[][]{bottke2016} - with more recent estimates usually proposing younger surface ages \citep[see][ for a full review]{Herrick2023}. A first popular resurfacing scenario suggests that Venus features an episodic lid tectonic regime, where one or several catastrophic lithospheric overturn events are followed by long-lasting quiescent episodes with limited geological activity (e.g. \citet[][]{Turcotte1993}; \citet[][]{ArmannTackley2012}). At present day, Venus would be in such a quiescent 'stagnant lid' period. Variations of this scenario include a current transition from mobile lid to stagnant lid regime (\citet[][]{WellerKiefer2020}) or a transition that occurred in the past (\citet[][]{WayDelGenio2020}). Recent work has also linked the convection regime transition to changes in surface conditions such as surface temperature \citep[][]{noack2012,gillmann2014}.

The alternative scenario, equilibrium resurfacing \citep[for example][]{ORourke2014}, is equally consistent with the crater distribution, which cannot be distinguished from a random one, \citet[][]{phillips1992}, and surface age estimates. Equilibrium resurfacing could occur in a stagnant lid-like regime \citep[][]{ORourkeKorenaga2015} with constant regional-scale resurfacing at a limited rate, possibly due to a plutonic squishy lid - a weak crust due to magmatic intrusions \cite[][]{Lourenco2020,Smrekar2023}. While it appears clear that Venus' mantle convection regime has changed over time, the precise succession of events and transitions is still debated: for a list of suggested tectonic evolution scenarios and more details about mantle dynamics, see also \citet[][]{Rolf2022}.\\

\subsubsection{Volcanoes and volcanic features}
In addition to the extensive basaltic plains, Venus exhibits tens of thousands of volcanoes \citep[][]{hahn2023} and volcanic features , such as shield-like volcanoes, cones, pancake domes, and many lava flows. The lava flows discovered in the Magellan radar data, among them Baltis Vallis being the longest known in the solar system with $>$6000~km length \citep[][]{Komatsu1992,conrad2023}, point towards abundant volcanic activity in the past. Theoretically, the higher density of the atmosphere would imply shorter lava flows than on Earth due to more efficient cooling of the material. However, taking into account the higher atmosphere and ground temperature, lower gravity, atmosphere heat capacity, winds, and the strong Infra-Red absoption by the CO$_2$ atmosphere could result in slow-cooling extended lava flows, up to 75\% longer than on Earth \citep[][]{flynn2023,snyder2002}.

The diverse volcanic features have also been associated with widely diverse lava viscosity respectively, and possibly different compositions. As such, long lava flows have been suggested to have a low viscosity and the lava composition to be possibly carbonatitic \citep{kargel1994}. Other features that have been proposed to be indicative of non-basaltic composition are steep sided domes and sinuous rilles and canali \citep[see][for a review on surface composition]{gilmore2023}. 

The above-mentioned $>$6000~km long Baltis Vallis lava channel was deformed since its formation \citep[][]{StewartHead1999};\citet[][]{StewartHead2000}. \citet[][]{McGregor2023} compared the long wavelength deformation of this lava flow, likely related to mantle convection processes, to the topography obtained from 3D spherical mantle convection simulations to estimate the viscosity of the Venus mantle. Their best-fitting models suggest Venus' mantle viscosity values are about two orders of magnitude smaller than Earth's. Theoretical calculations show that with improved gravity measurements to be expected from future Venus missions like EnVision, VERITAS and DaVINCI it will be possible to test this result, since the gravity response to periodic atmospheric loading can be used to obtain first-order estimations of both the mantle and core viscosity \citep[][]{Dumoulin2017, petricca2022}.

\subsection{Venus: active or dormant?}
\begin{figure*}
\centering
  \includegraphics[width=1.0\textwidth]{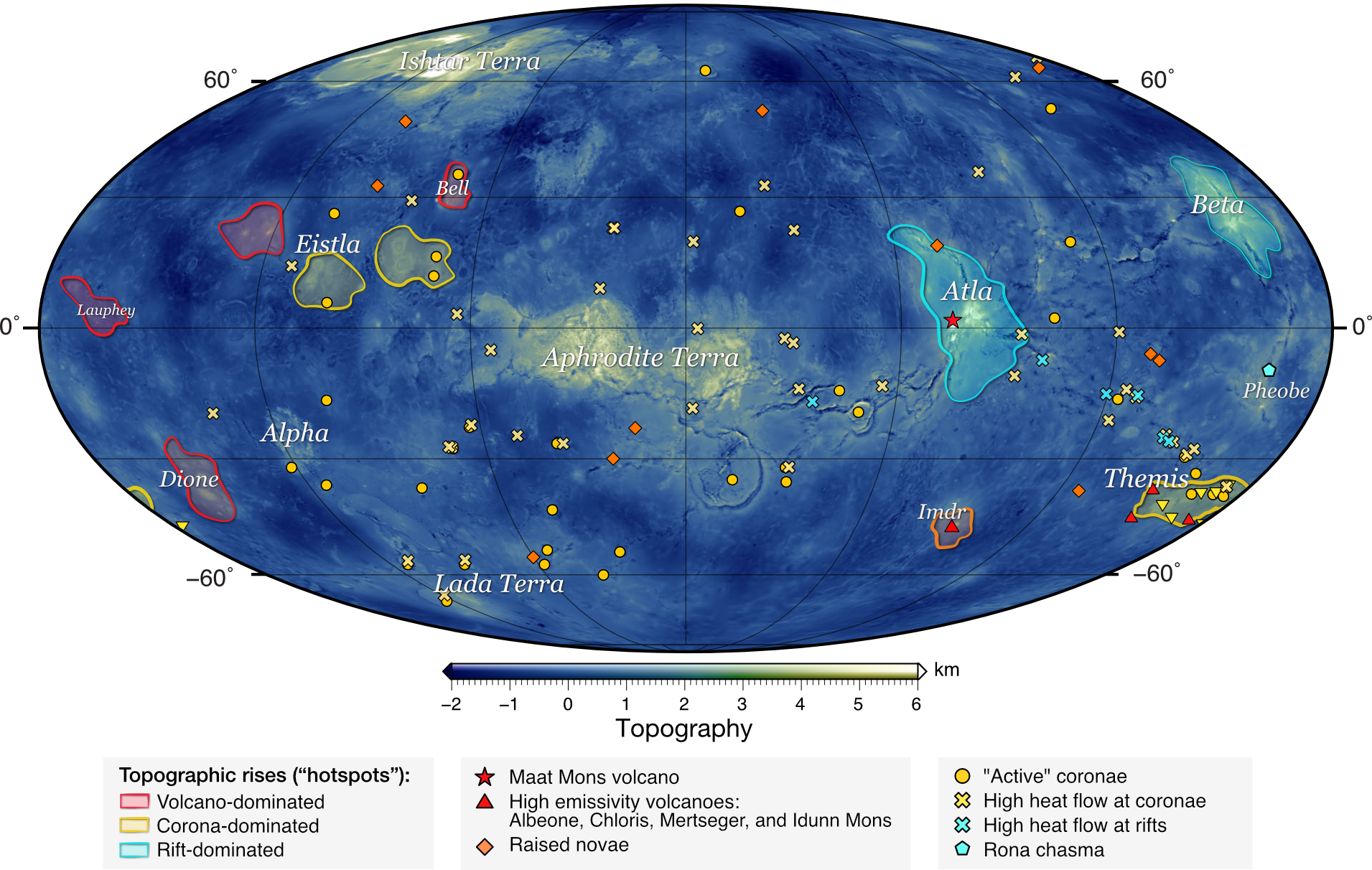}
\caption{Venusian sites of proposed recent magmatic/volcanic activity. Mollweide projection of Venus' global topography relative to 6,051.877~km \citep[][]{Ford1992}, centred at 120$^{\circ}$E longitude. The nomenclature of key areas is shown in white text. Different types of ``hotspot" locations: corona-dominated hostpots (yellow regions), rift-dominated hotspots (light blue regions), and volcano-dominated hotspots (red regions), from \citep[][]{stofan1995,smrekar1999,Smrekar2010}. Maat Moons is plotted as a red star \citep[e.g.,][]{Robinson1993,Herrick2023b}, whereas youthful volcanoes, as evidenced by their high thermal emissivity values, are plotted as red triangles \citep[e.g.,][]{Smrekar2010,Stofan2016}. Recent lava flows have been detected at Idunn Monns in the Imdr Regio \cite[][]{d2017idunn,Dincecco2021}. Raised novae structures (locations from \citet[][]{krassilnikov2003}) are shown as orange diamonds and have been proposed to be short-lived, young structures \citep[][]{Gerya2014}. Coronae in the Themis Regione with high thermal emmissivity values are plotted as yellow triangles \citep[][]{Smrekar2010,Stofan2016}. The yellow circles represent large coronae that were suggested to link to plume activity based on their topographic signatures \citep[][]{Gulcher2020}. Individual sites of high heat flow ($>$75~mW~m$^{-2}$) at coronae and rifts are plotted as x-symbols \citep[from][]{ORourke2018b,Russel2021,Smrekar2023}. Finally, radar-bright deposits in eastern Eistla, western Eistla, Phoebe, and Dione Regiones were suggested to reflect the early stage of recently renewed magmatic activity, particularly at the Rona Chasma (plotted as a blue pentagon, \citet{Campbell2017}). The perceptually uniform scientific colour map ‘Davos’ \citep[][]{Crameri2021} was used for this figure to prevent visual distortion of the topography data. Figure credit: A. Gülcher}
\label{fig:VenusActivity}       
\end{figure*}

The original belief that Venus is volcanically dormant, indicative of a current 'stagnant lid' phase, is increasingly challenged by numerous studies suggesting an active present-day Venusian surface. 
Increased SO$_2$ concentrations in the atmosphere have long been used as an argument for volcanic activity \citep[][]{Esposito1984}. The youth of several volcanoes and so-called "coronae" (discussed below) is corroborated by high emissivity values observed by Magellan \citep{Klose1992,Robinson1993}. In more recent years, observations by the ESA Venus Express Visible and Infrared Thermal Imaging Spectrometer (VIRTIS) revealed infrared excess (high emissivity) at several volcanic and coronae features \citep[][]{Smrekar2010,Stofan2016,d2017idunn}, adding further evidence for recent geological activity (see Figure \ref{fig:VenusActivity}). Individual lava flows being just a few years old were proposed at Idunn Monns (Imdr Regio, \citet[][]{d2017idunn}). Perhaps surprisingly, the strongest indication for present-day volcanic activity on Venus is based on synthetic aperture radar data obtained between 1990 and 1992 by the Magellan spacecraft. Comparing volcanic areas that were imaged multiple times by Magellan, an additional volcanic flow was identified on the side of the tallest volcano on Venus, Maat Mons (see Figure \ref{fig:VenusActivity}), indicating volcanic activity within the 8-months cycle gap in 1991 \citep{Herrick2023b}. 
Additionally, various estimates place the global volcanic production rates roughly between 0.1 and 10~km$^3$~yr$^{-1}$ \citep[for details, see][and citations therein]{Marcq2023}.
These recent lines of evidence strongly indicate that Venus is currently volcanically active, raising expectations of measurements of gas plume compositions by missions in the near future. 

Several hundreds of so-called "coronae", quasi-circular tectono-volcanic structures, scar the surface of Venus. Although their origin is still debated, coronae are often suggested to be formed by thermal plumes impinging and (partially) penetrating into the lithosphere, with different corona topographies representing different stages in the evolution \citep[][]{SmrekarStofan1997, Gulcher2020}. Based on the morphological differences between active and inactive coronae structures in geodynamical models and subsequent analysis of Venus observational data, numerous corona sites were suggested to host ongoing plume-lithosphere interactions underneath (\cite[][]{Gulcher2020}, see Figure \ref{fig:VenusActivity}). Numerical studies also led to the interpretation that raised novae (structures with a stellate fracture pattern) represent short-lived, geologically-active features \citep[][]{Gerya2014}. Artemis, a uniquely large corona feature on Venus, is suggested by numerous studies to host a retreating subduction zone on its southeastern rim. This argument has been based on observations \citep[e.g.,][]{SandwellSchubert1992}, numerical modelling \citep[][]{Ueda2008,Gulcher2020,Gulcher2023}, and analogue laboratory experiments \citep[][]{davaille2017}. While the south-eastern rim of the Artemis feature is perhaps the clearest surface manifestation of subduction on Venus, numerous other sites are also suggestive of ongoing regional subduction on Venus \citep[e.g.,][]{SandwellSchubert1992}. 

\subsection{Tectonic features and regimes}
All the evidence leaves little doubt that Venus has been very active (volcanically and tectonically) in the past and is likely still active at present-day. Yet, the precise nature of this activity remains uncertain. Venus tectonics shares some characteristics with the martian stagnant lid, but its surface is also marked by many traces of mostly local deformation and spatial variation (see tectonic features detailed by \citet[][]{solomon1992} and \citet[][]{mcgill2010}). Additionally, it is unlikely that a stagnant lid could remove enough heat from the planet's interior without observable widespread melting, i.e. an order of magnitude above the Earth's \citep[][]{ArmannTackley2012}.

Venus does not exhibit evidence of global Earth-like plate tectonics \citep[see][]{smrekar2018,Rolf2022}. Its surface does feature huge rift structures ("chasmata"), some of which may be sites of active extension, but lacks a global network of subduction, the primary counterpart to mid-ocean ridge extension on Earth. Locally, plume-induced subduction may be possible (see above), but it remains uncertain whether plumes are the sole driver of subduction initiation and extensional tectonics on Venus. Broad regions of plume-induced uplift ("hotspots" like on Earth) have been proposed (see Figure \ref{fig:VenusActivity}), but their link to regional tectonics and volcanism is diverse and perhaps non-unique \citep[][]{stofan1992,smrekar1999,Smrekar2010,Karimi2023}. 

Various work sheds light on the important role of lithospheric recycling other than subduction, such as lithospheric delamination, on surface deformation and horizontal displacements \citep[e.g.,][]{SmrekarStofan1997,Adams2022,Adams2023}. The basalt-to-eclogite phase change appears to drive gravitational instabilities and lithospheric delamination \citep{Adams2023, Gulcher2023}. Overall, Venus’ global tectonics and volcanism seem to be mainly driven by plumes, intrusive magmatism, and lithospheric delamination, which remains challenging to apprehend from a purely Earth-focused point-of-view. 

 Episodic activity, a transition of the tectonic regime or the plutonic squishy-lid (PSL) regime \citep[][]{Lourenco2020} could contribute to explaining the specifics of Venus. The plutonic squishy lid, in particular, has been one of the most interesting developments in modeling of the possible interior dynamics of Venus in the past decade. In the PSL, most of the magma generated by partial melting in the mantle does not reach the surface, but ends up as intrusions in the lithosphere. The PSL results in a thin lithosphere warmed locally by plutonism that provide rather diffuse weak zones, compared to Earth-like plate tectonics, conducive of deformation, without generating unrealistically large surface melt eruptions.   However, at present, we are not yet able to fully discriminate between these scenarios. As such, ``stagnant lid" is probably not the right term to use here, and ``single plate" planet may be preferable for now, as an intermediate between active and stagnant lid (\citet[][]{Rolf2022, Gillmann2022}). 

These differences, together with limited erosion, the lack of life and surface water, result in better-preserved geological records on Venus. Despite covering at best the last billion year, Venus could offer valuable insights into Earth's Archaean tectonic processes before plate tectonics operated on our planet \citep[][]{harris2013,hansen2018}. Venus also exemplifies the extreme variability in styles of dynamics and tectonics possible on Earth-sized terrestrial planets \citep[][]{way2023}.

\subsection{Surface processes and composition}
The surface of Venus does not only inform us about the dynamics of the interior. Information about chemistry and the interaction between the solid planet and the atmosphere can also be derived from its observation. We note that not all of Venus' surface is covered by volcanic material. Sedimentary processes have also been observed on Venus and most of the sedimentary material is produced by impacts on present-day Venus \citep[see][for a full review]{carter2023}. 

Pyroclastic volcanic activity is also a possible source for fine material \citep[][]{airey2015}, but this type of volcanism appears to be relatively rare \citep[][]{Campbell_Clark2006,Ghail_Wilson2015,Grosfils2000,Grosfils2011,Keddie1995,McGill2000}, due to Venus' dense atmosphere \citep[][]{Head1986} and the possibly low volatile content of its lava. Nevertheless, an examination of radar-bright deposits with diffuse margins using Magellan data led to the interpretation of pyroclastic flow deposits at various locations on Venus that indicate recently renewed magmatic activity, contributing to local uplift and extensional tectonism \citet[][]{Campbell2017}.

Fine-grained material has been observed to form dunes \citep[][]{greeley1992} or landslides \citep[][]{malin1992} under the action of aeolian erosion and transport, especially in the highlands \citep[][]{selivanov1982}. The morphology of some valleys has also been compared to patterns resulting from fluvial erosion \citep[][]{khawja2020}. However, identification of sedimentary rocks is still difficult and none have been observed on Venus with reasonable certainty, as other origins (like lava flows) could equally explain the observation, although candidates have been proposed \citep[][]{byrne2021}, and measured densities could suggest their presence.

Chemical weathering is also possible on Venus, but its effects still have to be quantified \citep[][]{dyar2021,teffeteller2022,carter2023}. They may be a secondary contribution to sediment production, but play a major role in surface mineralogy \citep[see][for a complete review]{gilmore2023}. Surface rocks can react with atmosphere components, such as oxidized species that make up a large part of the atmosphere, but also traces components like CO, or S-, Cl-, F-bearing molecules.

Chemical equilibrium between the atmosphere and the surface has been suggested to control the abundances of several species in the atmosphere of Venus, acting as a buffer \citep[][]{lewis1970venus,volkov1986}. Those species include traces components,like HCl and HF, but also major constituents, like CO$_2$ \citep[][]{fegley2014venus}. Buffering reactions remain uncertain to this day \citep[see][for more details]{zolotov2018,Gillmann2022}. Buffering of S species seems possible, but the buffering mineral is not obvious \citep[][]{zolotov2018}. HCl and HF buffers are more likely, involving chlorine- and fluorine-bearing minerals \citep[][]{fegley1997}. Buffering of water is unlikely due to the instability of hydrated material at the surface. The case of CO$_2$ depends on the availability of carbonates, which appear unstable under present-day surface conditions and the presence of SO$_2$, and the absence of water \citep[][]{zolotov2018}. 

On the other hand, surface conditions on Venus currently match the hematite-magnetite, hematite-magnetite-pyrite and magnetite-pyrite mineral equilibrium. If the atmosphere is buffered by surface mineralogy, an equilibrium like hematite-magnetite-pyrite could set the ratios of involved oxidized/reduced couples: CO/CO$_2$, SO$_2$/(COS, H$_2$S, S$_2$), and H$_2$/H$_2$O \citep[][]{zolotov2018}. However, the presence of hematite and magnetite could be the weathering product of fresh volcanic material exposed to oxidized species \citep[i.e.][]{gillmann2020,warren2022}. Indeed, recent work \citep[][]{filiberto2020,berger2019} has shown that oxidation processes involving oxidized species (CO$_2$, SO$_2$, H$_2$O, possibly O$_2$ in the past of Venus) with Fe-bearing material at the surface, such as fresh lava flows, to produce hematite and magnetite could occur on short timescales (within days). Therefore, some observed lava flows on Venus have been argued to be no more than several years old \citep[i.e.][]{Filiberto2014,filiberto2020}. 

Surface oxidation reactions could have implications for the fate of water loss on Venus, which are discussed in Section \ref{sec:evol}, further down, and in more detail in \citet[][]{Gillmann2022}. Laboratory work is a key factor in the understanding of the composition and evolution of the surface of Venus, however, the surface pressure-temperature range complicates the implementation of such experiments. The Glenn Extreme Environment Rig (GEER) at the NASA Glenn Research Center is an 800~L chamber that can operate at surface conditions with realistic gas composition (CO$_2$, N$_2$, SO$_2$, OCS, H$_2$O, CO, H$_2$S, HCl, HF) for tens of days. The first results \citep[][]{Rado22} showed the high reactivity of Ca-bearing pyroxenes and volcanic glasses, contrary to olivine and labradorite. The versatility of the chamber would help to improve the knowledge of solid-gas interactions at the surface. 

The Magellan radar properties highlighted strong differences for Maxwell Montes (65$^{\circ}$N) and Ovda Regio (and other highlands in the tropics), and therefore point out substantial differences in surface composition \citep[][]{Trei16}. On both sites, there is an increase in radar backscatter coefficient (radar brightness) from $\sim$0.05 in the plains up to $\sim$0.15 at 4~km above mean radius. However at higher elevations the behavior diverges totally. In Ovda Regio and other equatorial highlands, the radar backscatter coefficient increases with increasing altitude to reach typical values of 1, and then drops suddenly to values around 0.1 and lower. On the other hand, the radar backscatter coefficient and emissivity on Maxwell Montes continue to increase with a similar slope up to 5~km. Above this altitude, there is a strong increase of the radar backscatter up to around 2, with a respective decrease of the emissivity from 0.8 to 0.5. This transition is sometimes called the 'snow line'. The elevation of this snow line is not a constant elevation, several km higher in the North-West than in the South-East. The atmospheric composition may therefore be not constant across Maxwell, and could be indicative of depletion of its snow-producing component due to winds from the South-East \citep[][]{Stre22}. There are only few constraints on the composition of the surface of Maxwell Montes, only that it should be a mineral, or minerals, with a high dielectric constant. Bi/Te/S mixture \citep[][]{Port20} and Hg minerals \citep[][]{Port21} have been tested in the laboratory at Venus surface conditions. The later is unlikely due to the amount of mercury needed, whereas the first mixture is an encouraging candidate.

\subsection{Perspectives}
Many uncertainties regarding the surface and interior of Venus remain. Upcoming observations by a new generation of spacecraft will improve our understanding considerably. Since the viscosities of molten and solid iron cores vary by many orders of magnitude (\citet[][]{deWijs1998}; \citet[][]{Rubie2003}; \citet[][]{VanOrman2004}; \citet[][]{YunkerVanOrman2007}), it will be possible to distinguish whether the core of Venus is already solid or still at least some part of it is still molten. Clues regarding the past tectonic regime of the planet will also be gathered: we can expect that in case the planet experienced mobile lid tectonics for a significant part of its evolution, as on Earth, the more efficient cooling of the interior, compared to a stagnant lid, would have allowed for the formation of a solid inner core \citep[][]{ONeill2021}. Future seismological observations could constrain the core radius and inform us about the present-day state of the Venus core. In addition, the seismic velocity structure of the core could give us valuable clues regarding the composition of the core, thus allowing us to test various models suggested for the formation and early evolution of Venus. In parallel to this, near-infrared measurements from DAVINCI, EnVision and VERITAS would provide critical information on the mineralogy of the surface of Venus, possibly leading to valuable clues on mechanisms and exchanges throughout the history of Venus, and constraints on its past surface conditions, see also \citep{Gillmann2022}.

\section{The evolution of Venus and modeling}
\label{sec:evol}

The past history of Venus remains one of the biggest puzzles offered by Earth's closest relative. We still only have a partial understanding of the evolution of its internal, atmospheric or surface conditions. As discussed above, we possess very few constraints on the past or present state of the interior of Venus. The surface of Venus mostly yields data regarding the present-day or recent history. Only the Venusian atmosphere has preserved traces of earlier events, for example through isotopic ratios and elemental abundances. However, what evidence remains in the atmosphere is degraded because it is the result of 4.5 billion years of accumulated processes as varied as asteroid impacts, mantle outgassing, chemical reactions and atmospheric loss processes, among others. As a consequence, no observation can be attributed to only a single clearly identifiable cause mechanism. The evolution of Venus has to be considered as a combination of coupled processes and feedback and rarely allows a unique explanation. 

\subsection{Modelling interior-atmosphere feedback mechanisms}
The reader is suggested to check \citet[][]{Gillmann2022} for a description of the many mechanisms that affect the atmosphere of Venus over its evolution, and their consequences for surface conditions. \citet[][]{Rolf2022} provides the same effort concentrating onto the interior state and dynamics. The question of the hypothetical habitability of Venus is discussed in detail in \citet[][]{westall2023habitability}, while \citet[][]{salvador2023} provides an overview of the primordial evolution and magma ocean phase during Venus' first few million years. The point of view of comparative evolution of the terrestrial planets in the Solar System and why Earth, Mars and Venus diverged during their evolution is also fully described in \citet[][]{Hamano2023}. The synergies between Venus science and exoplanetary exploration is discussed in \citet[][]{way2023}. In the present work, we will only summarize the current state of the research beyond what has already been discussed above.

One of the main recent advances on the topic comes from the rise of so-called coupled numerical simulations (even though the importance of this approach had been highlighted long before, i.e. by \citet[][]{ingersoll1969runaway,walker1975evolution}). They generally focus on the planetary scale and the long-term (billion years) evolution. These simulations attempt to unite processes from various disciplines into a single self-consistent model in order to avoid prescribed transitions. 

\begin{figure*}
\centering
  \includegraphics[width=1.0\textwidth]{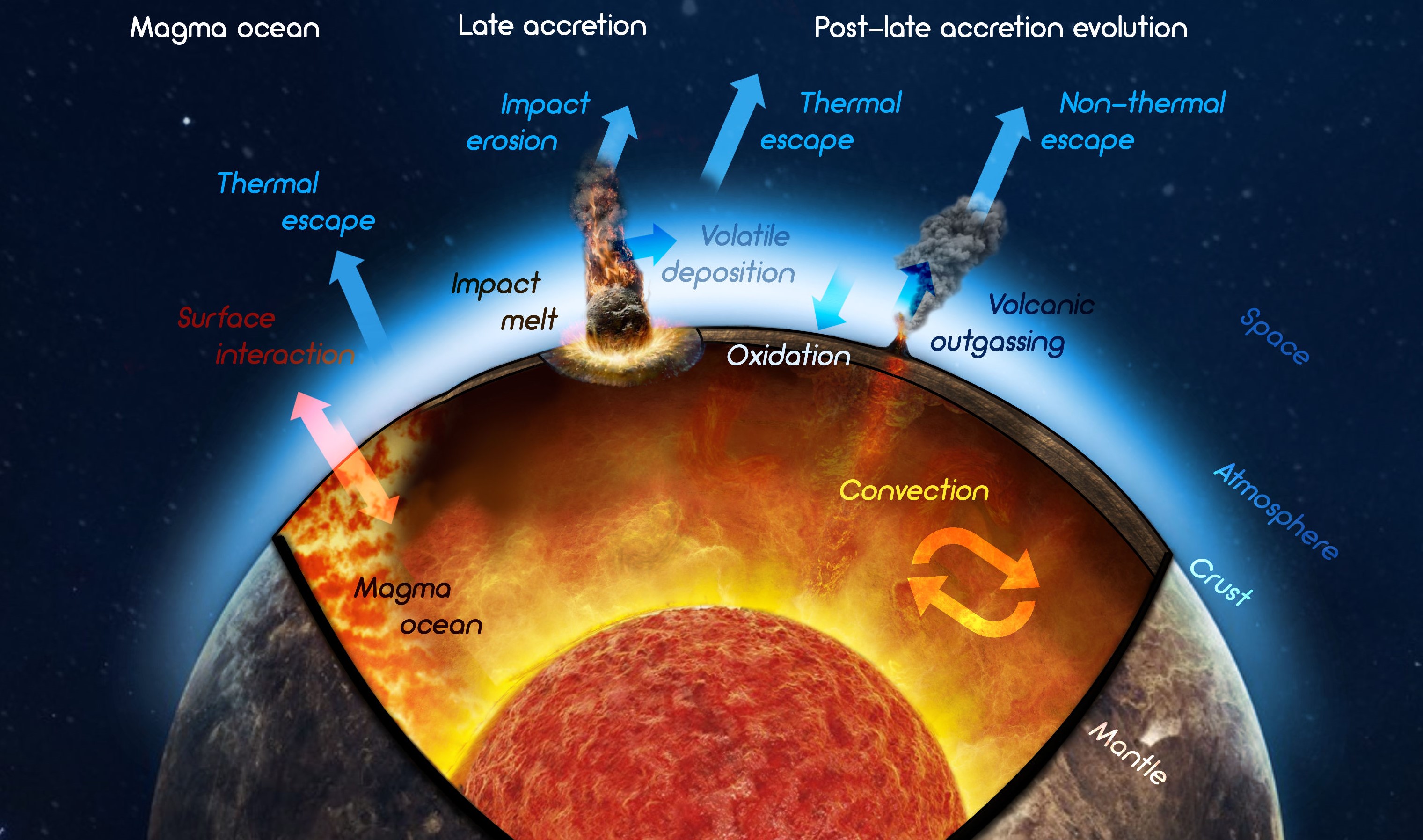}
\caption{Depiction of the volatile exchanges that affect the atmosphere and surface conditions throughout the evolution of Venus. Time evolution shows early epochs on the left and present-day on the right. Adapted from \citet[][]{gillmann2020}.}
\label{fig:volat}       
\end{figure*}

Modelled processes usually include a combination of mantle and core dynamics, volcanic production, physics and chemistry of mantle degassing, atmospheric escape, surface-atmosphere interaction, atmospheric radiative equilibrium and impact processes (see Figure \ref{fig:volat}). The output of the numerical simulations is compared to available observables, such as atmosphere composition, geophysical properties derived from surface observation, isotopic ratios and elemental abundances.

Reproducing plausible mantle dynamics for Venus with numerical models is still an ongoing endeavor and several scenarios have been proposed \citep[][]{Rolf2022}. Simulations try to reproduce as many ancillary features observed on Venus as possible: the overall young surface of Venus, estimated volcanic production rates, $^{40}$Ar atmospheric concentrations, crustal/lithosphere thickness, surface heat flow and topography. Many of the suggested scenarios include transitions between regimes of active lid convection (akin to plate tectonics) and stagnant lid-like behavior. Models have investigated possible origins for these transitions, ranging from a simple result of the secular cooling of the planet \citep[][]{weller2020} to changes in surface conditions. For example, \citet[][]{lenardic2008} and \citet[][]{gillmann2014} have proposed that low surface temperatures favored a plate tectonic-like behavior, while a warmer surface like Venus' present-day conditions were conducive of a stagnant lid. In the model, low temperatures imply a higher upper mantle viscosity that translates into higher convective stresses in the lithosphere. The lithosphere can be broken and surface mobility increases. On the other hand, \citet[][]{noack2012} have suggested the opposite: high surface temperatures reduce the viscosity contrast across the lithosphere and promote surface mobility. The two approaches use different atmosphere and rheology treatments. \citet[][]{gillmann2014} used a self-consistent coupled grey atmosphere radiative-convective model, models of atmospheric escape based on various mechanisms and mantle rheology including diffusion creep and plasticity, with mineral phase transitions in the convection code StagYY \citep[][]{ArmannTackley2012}. \citet[][]{noack2012} calculated surface temperatures based on a parameterization of results by \citet[][]{bullock2001} that included cloud physics, and further used a bulk exponential decay of volatile content of the atmosphere to parameterize the effect of escape mechanisms. They used mantle rheologies based on diffusion and dislocation creep without plasticity, with the Gaia mantle convection code \citep[][]{Huttig2008}. In both cases, the mobilization of surface with surface temperature changes is due to the response of the lithosphere to changes in stress, but the mechanism depends on the chosen rheology: plasticity vs dislocation creep \citep[for further discussion, see][]{gillmann2014}. Investigation of the effects of the rheology of the crust and upper mantle appears to be critical if we want to better understand the dynamics and tectonics of Venus \citep[][]{tian2023}.

As one essential factor that affects the atmosphere composition, the greenhouse effect and the resulting surface conditions, outgassing of volatiles has become a hot topic for long-term planetary modelling. The convection regime affects melting and the resulting outgassing. It has been suggested that a stagnant-lid could result in low outgassing rates due to the thick lid \citep[i.e.][]{dehant2019,guimond2021,weller2022}. However, a present-day Venus in a stagnant lid has been shown to lose heat inefficiently and could heat up enough to reach a much higher melt production than even present-day Earth \citep[][]{solomon1982,ArmannTackley2012}, which would affect volatile transport. Intrusive melt emplacement is possible, but would affect the heat transport \citep[][]{Lourenco2020}.

\subsection{Modelling outgassing}
Volcanism and outgassing have proven to be tricky to model, not only due to the lack of constraints, but also due to the complexity of the process, which involves mantle dynamics, melting, partitioning of volatiles into the melt, melt ascent through the lithosphere and chemical speciation at the surface \citep[][]{Gillmann2022}. State-of-the-art estimates use the planet's thermal and dynamic evolution to calculate mantle partial melting and volcanic production, either through 1-D parameterized \citep[][]{krissansen2021,driscoll2013,phillips2001}, or 2-D/3-D self-consistent approaches \citep[for example][]{noack2012,gillmann2014,ortenzi2020}. Despite uncertainties, especially on initial conditions, these models remain the only self-consistent approach toward quantitative long-term volcanic evolution. 

Partitioning of volatile species into the silicate melt during partial melting of the mantle depends on the melt fraction and the chemical equilibria involving those species. Therefore, it is governed by the local pressure and temperature, as well as the mantle composition and redox state \citep[i.e.][]{pawley1992,hirschmann2008}.
Melt extraction modelling is often simplified, and it is assumed that a set fraction of magma production reaches the surface. Recent work \citep[][]{Lourenco2020} have refined modeling of the emplacement of melt by including the treatment of intrusive magmatism, leading to a warm lithosphere with localized melt inclusions (the plutonic squishy lid discussed above). 

The physics and chemistry of outgassing itself have also been improved in simulations over the past ten years. Water solubility in silicate melts has long been known to increase with pressure \citep[for example][]{tuttle1958,moore1998}, but \citet[][]{gaillard2014} have reminded modellers of the importance of that fact \citep[see][for a review]{Marcq2023,Gillmann2022}. In summary, the solubility in the silicate melt of various species that can be outgassed by volcanism varies with surface and magma conditions (pressure, temperature and oxygen fugacity). The combination of these conditions determines the stability of the species. In particular, while water is only moderately soluble at Earth's 1 bar surface pressure, it becomes very soluble at $\approx$100 bar. As a consequence, on Earth, outgassing is dominated by water (with significant SO$_2$ and CO$_2$), while at present-day on Venus, water might not significantly outgas, and CO$_2$ and CO would be the main volcanic products. Recent planetary evolution models have incorporated the influence of surface pressure on outgassing \citep[for example][]{ortenzi2020,weller2022}. Such models suggest that the dry atmosphere of Venus may not be incompatible with a wet interior \citep[see also][]{Gillmann2022}. A wet lithosphere, despite high surface temperatures, has been suggested to be detectable in the future, using seismic velocity profiles, as hydrous minerals could be conducive of distinctly lower velocities than anhydrous ones \citep[][]{chen2024}. Low water outgassing may make detection of volcanic plumes in a CO$_2$ dominated atmosphere challenging, but manageable, as plumes that are depleted in water relative to the background atmosphere could possibly be observed \citep[][]{Marcq2023}. The outgassing of SO$_2$ may also be limited by high surface pressures, which will need to be reconciled with observed temporal variations of the SO$_2$ mixing ratio, suggested to be caused by volcanic activity \citep[e.g.][]{encrenaz2020}. Volcanic plumes may thus prove difficult to separate from background atmosphere using this species, especially for reduced mantle conditions \citep[][]{Marcq2023}. Moreover, from an evolutionary point of view, variations of outgassing with total atmospheric mass and redox state imply that it could be very difficult to estimate outgassing in isolation, without considering the coupling between the mantle and the atmosphere.

\subsection{Modelling volatile sinks}
Atmospheric sinks of volatiles are also being investigated and present-day observations are used to extrapolate possible past loss rates, assessing whether they could be consistent with the present-day D/H ratio. The first type of volatile sink is the atmospheric escape \citep[see][for a general review, and for Venus, respectively]{gronoff2020,Gillmann2022}. CO$_2$ does not appear to escape substantially. Water loss in the form of H$^+$ and O$^+$ has been measured on Venus with an average O$^+$ escape rate of (3-6)~$\times$~10$^{24}$~s$^{-1}$ and a H:O stoichiometry close to 2:1 \citep[][]{futaana2017}.The most recent estimates \citep[][]{Persson2018} are consistent with the lower range of values obtained, and highlight variations of the loss rates and H:O ratio over the solar cycle. Extrapolations into the past of Venus remain uncertain. Using the assumption of constant atmosphere structure and composition, calculations indicate that a global equivalent layer of a few centimeters to a few tens of centimeters of water could have been lost to space over the last 3.7~Ga \citep[][]{Persson2020}. Further advancements in the field of modelling accurate escape mechanisms for a wider range of conditions (composition, past EUV flux, magnetic field configurations) would be a huge benefit for comparative planetology, including efforts focusing on Venus.

Hydrodynamic (thermal) escape is thought to be much more efficient at removing volatiles from the atmosphere, especially water. An atmosphere dominated by water vapour that could be photo-dissociated is required for hydrodynamic escape to occur. Given enough energy sources, such as strong Extreme UV radiation or heating from the base by a magma ocean, the steam atmosphere could expand and escape. Therefore, hydrodynamic escape is generally limited to the first few hundred million years of the evolution of the planet or the magma ocean phase. Calculations suggest it could extract multiple Earth-oceans worth of hydrogen from the atmosphere in a few tens to hundreds of million years. This would leave behind large amounts of oxygen \citep[i.e.][]{hunten1973,KastingPollack1983,chassefiere1996,Gillmann2009}, as oxygen escapes less efficiently than the lighter hydrogen. Attempts have been made to constrain this with stable noble gas isotopic ratios. Scenarios that fit measured isotopic ratios can be reached, but no unique solution has been found due to the many uncertain parameters involved, like atmosphere composition and structure, efficiency of energy deposition or early extreme UV flux \citep[][]{Gillmann2009,Odert2018}. 

The left-over oxygen from the hydrodynamic escape phase, or later released, must either react, be lost or be trapped in order to reach the present-day state of the atmosphere. The need to remove oxygen from the atmosphere has been proposed as a constraint for models of the evolution of Venus \citep[][]{KastingPollack1983,gillmann2014,gillmann2020,warren2022} and was used to estimate the composition of the late accretion impactors and the late delivery of water, CO$_2$ and N$_2$ to Venus \citep[][]{gillmann2020}, concluding that after the main accretion phase most impactors could have been relatively dry akin to enstatite chondrite, as was found for Earth \citep[i.e.][]{dauphas2017}.

Atmospheric non-thermal escape has long been considered to be the main sink of oxygen during the evolution of Venus. The more recent escape measurements however, suggest that only a small amount of oxygen can actually be lost to space. As a result, other processes have been considered, such as gas-surface reactions. The oxidation of the solid surface and the iron-bearing material of fresh lava flows into hematite or magnetite was suggested to be a possible sink of oxygen \citep[][]{lecuyer2000}. It has been calculated that the efficiency of such a process is likely very low \citep[][]{gillmann2020,warren2022}, even if the reaction itself is very fast. Such reactions would quickly form a thin surface coating. The oxidation rate has been estimated at 30~$\upmu$m over 500,000~years \citep[][]{dyar2021,teffeteller2022}. In the end, solid surface oxidation is unlikely to be a more important sink than atmospheric escape.

Another possible way to trap oxygen could include oxidation of hot volcanic ashes \citep[][]{warren2022,warren2023}, which would be more efficient than the surface reaction due to their larger surface-to-volume ratio. However, Venus offers few examples of explosive volcanism. Finally, it has been long suggested that a magma ocean could become an important reservoir of oxygen, as vigorous convection could achieve conditions closer to equilibrium with the atmosphere \citep[][]{KastingPollack1983,Gillmann2009,schaefer2016,wordsworth2018,wordsworth2022,warren2023}. In case a long-lived magma ocean exists, it could trap the excess oxygen in the planet and solve the puzzle. However, recent modelling work on steam atmospheres suggests that for a Sun-like spectrum, the surface of Venus would be able to solidify before water is lost by hydrodynamic escape \citep[][]{selsis2023}. According to the models, the occurrence of a magma ocean could even be prevented under certain conditions. The low surface temperatures obtained in these models derive from the radiative component of the structure of steam atmospheres, rather than the fully convective structure often assumed in previous work. Such short-lived, or absent, magma oceans would have major implications for the evolution of Venus and remove a possible important sink for oxygen.

\subsection{Modelling the core}
A final recent topic of interest in modeling investigates the core thermal evolution and magnetic field generation. On terrestrial planets, self-sustaining magnetic fields are generated in the liquid core by convective movement (the core dynamo), which is governed by the thermal history of the planet. Thermal evolution depends on the mantle evolution and the efficiency of heat transfer out of the solid planet at the surface. Surface conditions are coupled with the mantle evolution, as discussed before. Finally, the magnetic field itself affects atmospheric escape in various ways and has long been assumed to have shielded the atmosphere from intense water loss, as evidenced by the lack of water on Venus and Mars compared to Earth, the only body with a significant self-generated magnetic field. However, recent observations suggesting that the loss of water at present-day was roughly similar on Earth, Mars and Venus challenged this point of view \citep[see][]{ramstad2021}. The question is still debated, with the only certainty being that magnetic fields have an effect on escape mechanisms, albeit a poorly understood one \citep[][]{Hamano2023,Gillmann2022,way2023}. The reasons why Venus has no magnetic are mostly addressed through modeling and have already been discussed in section \ref{sec:intern} \citep[][]{orourke2018a,ORourke2020,Jacobson2017,Dumoulin2017}.


\subsection{Plausible evolutionary pathways}
There is still no consensus about the evolution of Venus. Venus is generally thought to be in a post-runaway (or moist) greenhouse state \citep[][]{ingersoll1969runaway}. In an atmosphere above a liquid water ocean, as temperature increases - for example, due to an increase in absorbed solar radiation - water evaporates and water vapor abundance increases in the atmosphere. In a normal greenhouse state, the increase in temperature leads to higher emitted thermal radiation, in order to preserve the energy balance between incoming and outgoing radiation. Water vapor absorbs outgoing thermal emissions, which affects the equilibrium temperature. However, there is a tipping point. As water vapour concentration increases and temperature increases, the structure of the troposphere is set by moist convection and water vapour saturation. This fixes the altitude and temperature where outgoing emission can occur and thus the maximum outgoing radiation itself: increased temperatures do not result in higher emissions any longer. If absorbed incoming radiation exceeds this limit (which is largely determined by the albedo and clouds for a given orbit), runaway greenhouse occurs. Then the atmosphere temperature increases until about 1400~K and it can radiate in the near-infrared, where it is no longer absorbed by water vapour \citep[see for example][]{goldblatt2012}. The planet exits this state as water is lost to space and the atmosphere desiccates. The exact timing, causes and conditions of the transition between the surface condition states are where scenarios diverge. Observations and simulations have informed three main end-member evolutionary pathways that would fit the current constraints on Venus history \citep[see also][]{Gillmann2022,westall2023habitability,Hamano2023}. 

In the first scenario, Venus would have followed an Earth-like evolution early on, retaining enough water while cooling down (due to the high albedo from a thick cloud cover on the day-side) at the end of the magma ocean phase to presumably condense surface liquid water and start a carbon sink that would have removed the thick initial CO$_2$ and left the planet with a thinner Earth-like N$_2$ atmosphere \citep[][]{WayDelGenio2020}. Mild surface conditions could have made Venus habitable for an uncertain period of time, before an event destabilized the climate and pushed it into an extreme runaway greenhouse, causing the loss of water. CO$_2$ would then be released by destabilized carbonate deposits and volcanism, for example. Later still, volcanic resurfacing produced the currently observable Venusian state. The nature of the destabilization is not precisely known, but it originates presumably within the mantle and has been suggested to be akin to Earth's Large Igneous Provinces on a larger scale \citep[][]{way2022}. It is also possible that sufficiently large impacts could have played a role, especially during the first few hundred million years of the evolution of the planet \citep[][]{Gillmann2016,o2017}.

In a second evolution scenario, Venus would lose its water efficiently during the magma ocean phase, without any water condensation \citep[as a type II planet][]{hamano2013emergence,Gillmann2009}. Unlike short-lived magma oceans on Earth-like planets (type I), Venus' magma ocean would have solidified slowly enough for water to be lost to space by hydrodynamic escape without condensation, possibly due to clouds forming on the night-side \citep[][]{turbet2021day}, and CO$_2$ would accumulate and form a dense atmosphere, lacking significant sinks. Therefore, Venus would have had a thick CO$_2$ atmosphere for most of its evolution, with only moderate contribution by later volcanism.

In a third possible scenario, with an outcome similar to the second one, water would not be lost to space. Instead, outgassing from an oxidized magma ocean is limited by high surface pressures and its redox state. This would be caused by the early degassing of other low solubility species (CO$_2$ for example \citet[][]{gaillard2022}). In the stifled outgassing scenario, the interior of the planet may be water-rich, but H$_2$O never reaches the atmosphere in significant amounts.

We are currently unable to discriminate between these scenarios, as available observations that relate to the past of Venus are often ambiguous and can be interpreted in multiple ways (such as the D/H ratio, or radiogenic Argon, for example). There is also possible overlap between the scenarios, depending on uncertain parameters (such as early albedo or atmosphere composition; see \citet[][]{salvador2023}): for example in an intermediate evolution between scenario 1 and 2, it is conceivable that, from the remains of the magma ocean phase atmosphere, a short-lived thin surface liquid water layer could have formed in the first billion years of Venus' evolution before being lost. The three main scenarios described here illustrate extreme end-members, with considerable leeway for variations in-between. Nonetheless, these scenarios have identified the critical role of water and water exchanges, as a marker of past surface conditions and a key to reconstructing Venus' history. The upcoming missions are building on this fact to focus on observations that could favor a specific pathway.

\section{Future missions}

The past decade has seen the interest in Venus exploration rise anew in the planetary science community, as the planet's critical role as a counterpoint to Earth's evolution gained visibility. The gaps in our understanding of Venus become even more apparent in light of Mars science advances and the need for more detailed surveys increasingly pressing as more exoplanets are being discovered.
The incoming (selected and in preparation) missions aim at bringing us missing clues about Venus' past and evolution, in particular when it comes to the fate of water and surface conditions \citep[see][for a full review]{widemann2023}. Here, we quickly summarize the focus of each of the three missions in advanced planning stages at the time of the writing of this review: DAVINCI (NASA), VERITAS (NASA) and EnVision (ESA). Their overarching goal is to solve the question regarding the atmospheric origin and planetary evolution: What is the origin of Venus' atmosphere, and how has it evolved? Was there an early ocean on Venus, and, if so, when and where did it go? How and why is Venus different from (or similar to) Earth, Mars, and exo-Venuses? All three contemporary Venus-focused missions were selected on their own merits, but also have developed strong synergies regarding their overlapping science objectives, due to the long-standing and unanswered questions about Venus.

\subsection{DAVINCI}

The Deep Atmosphere Venus Investigation of Noble gases, Chemistry, and Imaging (DAVINCI) mission was selected by NASA in 2020 with an anticipated launch date in 2029 and Venus atmospheric entry in 2031. Prior to entry, DAVINCI will set up the correct geometry for probe drop-off with two gravity assist flybys that will occur in 2030. 

DAVINCI will address questions related to Venus’s origin, evolution, and present state and complement the two orbiter missions that feature next-generation radar and nightside near-infrared (NIR) emission instruments for mapping the surface.
DAVINCI is the only mission that will purposefully enter the atmosphere and also has imaging capability on its carrier spacecraft to address the following topics: (1) Atmospheric origin and planetary evolution, including the history of water, (2) Atmospheric composition and surface interaction, and (3) Surface properties. DAVINCI will tackle these topics by making direct measurements of the atmosphere and its composition, including the complete suite of noble gases and confirmation of the D/H ratio in water. Together, these constrain the history of outgassing and atmospheric loss on Venus. The definitive in-situ analyses of near-surface gases will reveal, for example, the interaction between the surface and deep atmosphere. DAVINCI's in-situ investigations and new observations of the topography and NIR reflectivity of a representative tessera will then be linked to test hypotheses of water–rock interactions and water loss over time.

DAVINCI will collect measurements throughout the entire atmospheric column. However, the atmospheric region of acute interest to geochemists is at the near-surface because of possible surface-atmosphere reactions. Little is known about this lowest atmospheric zone because previous atmospheric entry missions returned minimal data, either due to instrument design or failure \citep[][]{fegley2014venus,johnson2019venus}. DAVINCI is the mission that hopes to close this knowledge gap by collecting atmospheric data continuously as the descent sphere drops towards the surface.

\begin{table}[]
 \hspace{-2cm}
\begin{tabular}{@{}lll@{}}
\toprule
Instrument                                                                                                                           & Description                                                                                                                                                                                & Goals                                                                                                                                                                                                                              \\ \midrule
\multicolumn{3}{|c|}{Descent   Sphere Instruments}                                                                                                           \\ \cmidrule(r){1-3}
\multicolumn{1}{|l|}{\begin{tabular}[c]{@{}l@{}}VMS \\ (Venus Mass   \\ Spectrometer)\end{tabular}}                                  & \multicolumn{1}{l|}{\begin{tabular}[c]{@{}l@{}}Quadrupole mass \\ spectrometer\end{tabular}}                                                                                               & \multicolumn{1}{l|}{\begin{tabular}[c]{@{}l@{}}Measure a broad suite   \\ of trace gases, including noble gases\end{tabular}}                                                                                                      \\ \cmidrule(r){1-3}
\multicolumn{1}{|l|}{\begin{tabular}[c]{@{}l@{}}VTLS (Venus \\ Tunable Laser \\ Spectrometer)\end{tabular}}                          & \multicolumn{1}{l|}{\begin{tabular}[c]{@{}l@{}}Tunable laser spectrometer \\ with a multipass Herriott cell \\ and three laser channels at 2.64,   \\ 4.8, and 7.4 mm\end{tabular}}        & \multicolumn{1}{l|}{\begin{tabular}[c]{@{}l@{}}Targeted measurements of key   \\ gases containing H, S, C, and O, \\ and their isotopes, including \\ D/H   ratio\end{tabular}}                                                       \\ \cmidrule(r){1-3}
\multicolumn{1}{|l|}{\begin{tabular}[c]{@{}l@{}}VenDI (Venus \\ Descent   Imager)\end{tabular}}                                      & \multicolumn{1}{l|}{\begin{tabular}[c]{@{}l@{}}NIR imaging camera with \\ broad band (720-1040~nm) \\ and narrow band (980-1030~nm)\end{tabular}}                                          & \multicolumn{1}{l|}{\begin{tabular}[c]{@{}l@{}}Measure surface topography and \\ constrain composition of the surface \\ at Alpha Regio during   probe descent\end{tabular}}                                                      \\ \cmidrule(r){1-3}
\multicolumn{1}{|l|}{\begin{tabular}[c]{@{}l@{}}VASI (Venus \\ Atmospheric\\ Structure \\ Investigation)\end{tabular}}                  & \multicolumn{1}{l|}{\begin{tabular}[c]{@{}l@{}}Sensors to measure  \\  pressure, temperature, \\ and dynamics\end{tabular}}                                                                & \multicolumn{1}{l|}{\begin{tabular}[c]{@{}l@{}}Reconstruction of the descent \\ profile and thermodynamic context \\ for other measurements\end{tabular}}                                                                          \\ \cmidrule(r){1-3}
\multicolumn{1}{|l|}{\begin{tabular}[c]{@{}l@{}}VfOx (Venus \\ Oxygen Fugacity \\ Experiment)\end{tabular}}                          & \multicolumn{1}{l|}{\begin{tabular}[c]{@{}l@{}}Solid-state Nernstian   \\ ceramic oxygen sensor;\\ a student collaboration   \\ instrument involving \\ Johns Hopkins University\end{tabular}} & \multicolumn{1}{l|}{\begin{tabular}[c]{@{}l@{}}Measure oxygen fugacity \\ in the near-surface environment\end{tabular}}                                                                                                            \\ \cmidrule(r){1-3}
\multicolumn{3}{|c|}{Spacecraft   Instruments}                                                                                                                                                                                                                                                                                                                                                                                                                                                                                               \\ \cmidrule(r){1-3}
\multicolumn{1}{|l|}{\begin{tabular}[c]{@{}l@{}}VISOR (Venus \\ Imaging System for \\ Observational \\ Reconnaissance)\end{tabular}} & \multicolumn{1}{l|}{\begin{tabular}[c]{@{}l@{}}One UV camera (755-375~nm) \\ and three NIR cameras\\  (930-938, 947-964, 990-1030~nm)\end{tabular}}                                        & \multicolumn{1}{l|}{\begin{tabular}[c]{@{}l@{}}Measure cloud morphology and \\ motion in the UV; measure surface \\ thermal emission with   correction for \\ scattered light and variable cloud \\ opacity in the NIR\end{tabular}}  \\ \cmidrule(r){1-3}
\multicolumn{1}{|l|}{\begin{tabular}[c]{@{}l@{}}CUVIS (Compact \\ Ultraviolet to Visible \\ Spectrometer)\end{tabular}}                                    & \multicolumn{1}{|l|}{\begin{tabular}[c]{@{}l@{}}High resolution UV   \\ spectroscopy and hyperspectral \\ imaging. A technology\\  demonstration instrument\end{tabular}}                                         & \multicolumn{1}{|l|}{\begin{tabular}[c]{@{}l@{}}Measure upper cloud   \\ composition, with a focus on \\ the unknown UV absorber\end{tabular}}                                                                                                         \\ \bottomrule
\end{tabular}
\caption{The DAVINCI instruments work together to comprehensively understand Venus.}
\label{tab:davinci}
\end{table}

\begin{figure*}
\centering
  \includegraphics[width=1.0\textwidth]{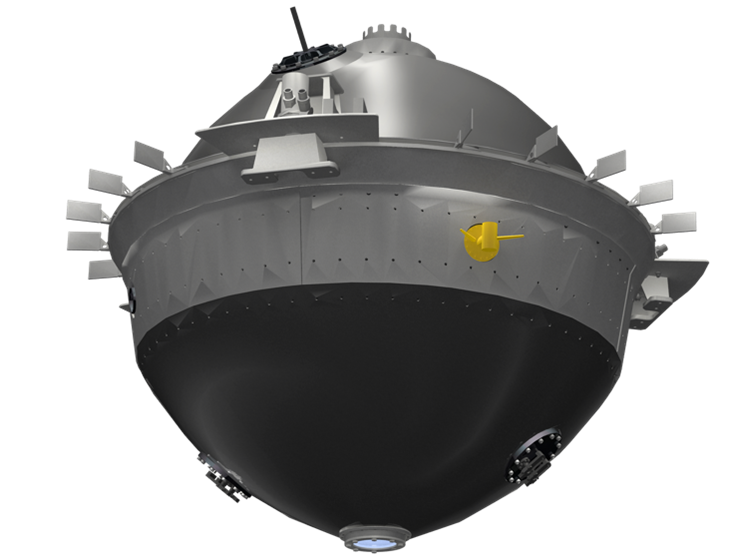}
\caption{The titanium DAVINCI descent sphere, approx. 1 m diameter,  protects its instruments from the extreme Venusian environment.}
\label{fig:Venusdavinci}       
\end{figure*}

The mission consists of a carrier relay imaging spacecraft with two instruments and the primary element of this mission: a descent sphere with five instruments (Figure \ref{fig:Venusdavinci}) that will be released into the atmosphere above Alpha Regio, an enigmatic tessera terrain. These instruments and their goals are summarized in Table \ref{tab:davinci}.

During the gravity assists, the carrier spacecraft will obtain flyby imaging at ultraviolet (UV) and near-infrared (NIR) wavelengths to gain new data about the clouds and poorly understood regions of the surface \citep[][]{garvin2022revealing}. VISOR (Venus Imaging System for Observational Reconnaissance) will measure surface near-infrared thermal emission from beneath the clouds on the planet’s night side, constraining surface composition at regional scales. In the UV on the planet’s day side, VISOR will measure cloud patterns and cloud motion. Meanwhile, CUVIS (Compact Ultraviolet to Visible Spectrometer) will provide detailed new measurements of the cloud tops to help constrain the identity of the unknown UV absorber. 

Probe descent will occur in June 2031. The probe will take approximately one hour to fall through the atmosphere, and this interval will be densely packed with carefully timed measurements, providing our most complete look into the composition and chemistry of Venus’ deep atmosphere, including trace and noble gases, key isotopes, and oxygen fugacity (Table 1).

By measuring the noble gas isotopes, Venus’s formation and evolution can be placed into context within the Solar System. Previously, only Ne and Ar were measured with robust confidence. Noble gases can serve as ‘atmospheric fossils’ and provide clues regarding Venus’s origin and evolution: for example, radiogenic noble gas isotopes can provide insights into volcanic outgassing over time. Xenon has also never been measured at Venus, and its depletion, relative to Earth and Mars is a diagnostic of giant impact processing and/or thermal escape. 

Another isotopic ratio of particular interest is D/H, often used to infer lost water over planetary evolution. DAVINCI’s vertically resolved new measurements at better accuracy and precision will aid in deconvolving Venus’ past and provide valuable constraints for models of water on Venus through time.

DAVINCI's measurements of chemically active gases will constrain coupled chemical processes and circulation of the sub-cloud atmosphere. Vertical composition profiles and gradients in the deep atmosphere are needed to constrain abundances of atmospheric volatiles, physical processes (e.g., circulation, predicted CO$_2$–N$_2$ gas separation \citep[][]{Lebo17}, thermochemical and photochemical reactions among gases (e.g., Krasnopolsky 2007, 2013), and the chemical interactions at the atmosphere–surface interface \citep[][]{fegley1997,zolotov2018}. For many of these reactions, the oxidation state is important but is not yet known. It will be derived using the VfOx sensor, which will directly measure the oxygen partial pressure and other clues from VMS and VTLS near-surface measurements of chemically reactive gases. The combination of gathering atmospheric compositional data simultaneously with temperature and pressure data by VASI, will not only provide a highly detailed snapshot of the atmosphere, but will also provide vital chemical data for modelling potential reactions and chemical cycles. 

Visual data of the enigmatic Alpha Regio region will be taken by VenDI through a sapphire window at the bottom of the descent sphere. The sequence of nadir-oriented near-infrared images will be initiated once the probe passes beneath the clouds and will last until surface touchdown. These images will provide new views of surface features at meter scales and potentially differentiate compositional patterns such as felsic rocks (possible markers of past water) that might exist at Alpha Regio \citep[][]{Gilmore2015}. Images from 1.5~km to the surface will feature a spatial resolution of less than 1 m, allowing erosional studies relating to the environmental history of Venus. Three-dimensional topography will be derived from structure-from-motion digital elevation maps. 

\subsection{VERITAS}

The Venus Emissivity, Radio Science, InSAR, Topography, and Spectroscopy (VERITAS) mission was selected by NASA as a Discovery class mission on June 2, 2021. It is designed to understand Venus’ evolution through the acquisition of foundational, high-resolution global datasets supported by two instruments and a gravity science investigation. 

\begin{itemize}
    \item The Venus Interferometric Synthetic Aperture Radar (VISAR) X-band datasets include: (1) a global digital elevation model (DEM) with 250~m postings and 6~m height accuracy, (2) Synthetic aperture radar (SAR) imaging at 30~m horizontal resolution globally, (3) SAR imaging at 15~m resolution for targeted areas, and (4) surface deformation from repeat pass interferometry (RPI) at 2~cm precision for targeted areas (Hensley et al., 2020).
    \item The Venus Emissivity Mapper (VEM; \citet[][]{helbert2022}) covers $\ge$80$\%$ of the surface in six NIR bands located within five atmospheric windows sensitive to iron mineralogy, plus eight atmospheric bands for calibration and water vapor measurements.
    \item VERITAS’s low circular orbit ($\le$ 250~km, mean 217~km) and Ka-band uplink and downlink to create a global gravity field with 3~mGal accuracy at 155~km resolution – a significantly higher and more uniform resolution than available from Magellan data).
\end{itemize}

\begin{figure*}
\centering
  \includegraphics[width=1.0\textwidth]{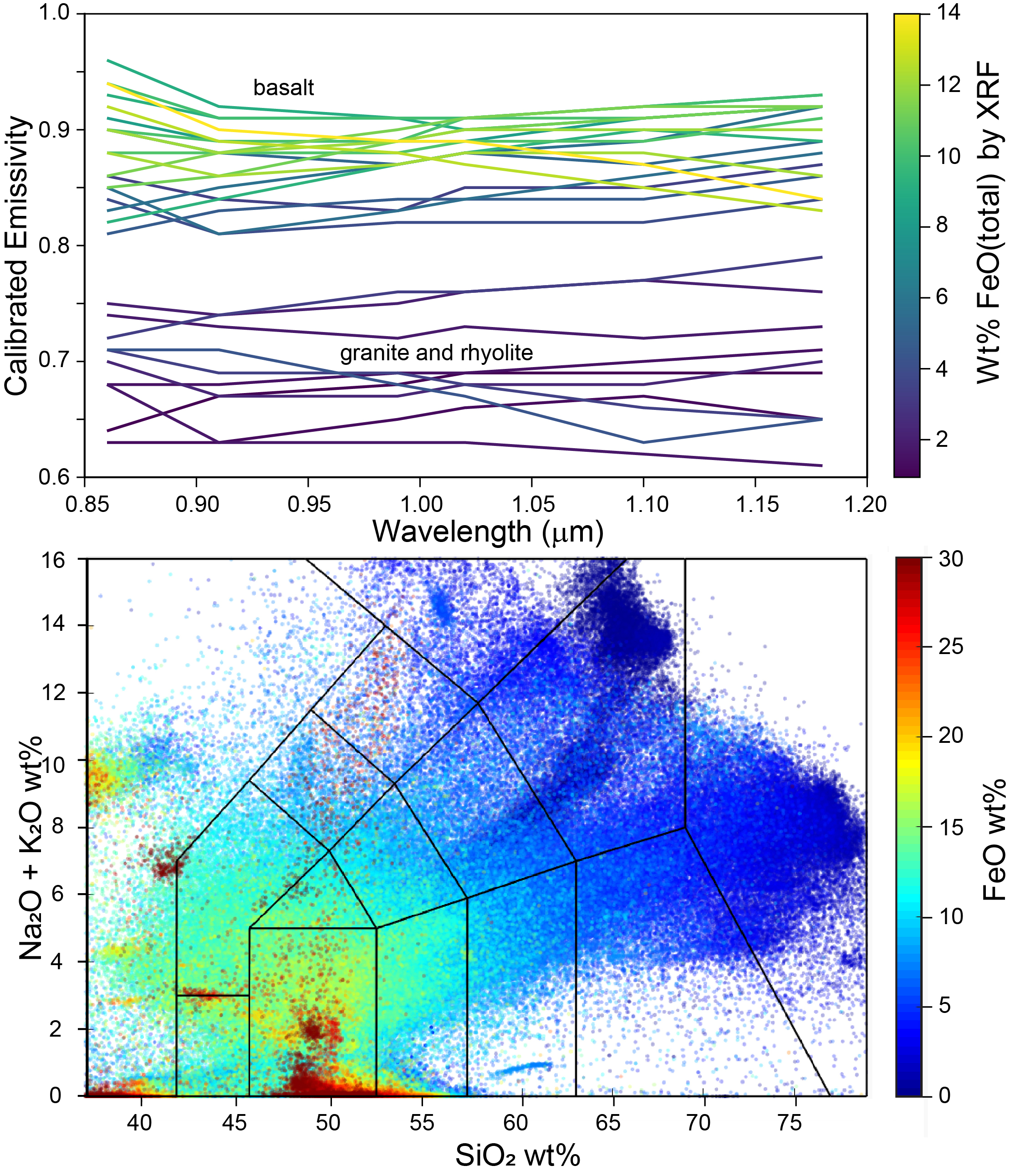}
\caption{(top) Laboratory measurements of emissivity at Venus temperatures enable basaltic and felsic rock types to be clearly distinguished due to variation with total Fe content (here expressed at total FeO). (bottom) Plot of total alkalis versus silica contoured with FeO content for volcanic rocks in the GeoRoc database; purple is high Fe and red is low (see key). Because of the inverse correlation between FeO and SiO$_2$, detection of FeO contents by VEM on VERITAS will allow distinctions to be made along the fractional crystallization trend between mafic and felsic rocks. Adapted from \citet[][]{dyar2020,dyar2021}}
\label{fig:VenusVeritas}       
\end{figure*}

In addition to their broad significance to geophysical and geological questions relating to Venus, all of these investigations will shed light on the geochemistry of Venus in different ways. A key emphasis is on the search for chemical fingerprints of past water in the tessera plateaus and larger inliers. Comparisons with calibrated emissivity measurements in the laboratory (Figure \ref{fig:VenusVeritas}) enable VEM to determine whether tesserae are globally felsic or mafic, and thus analogous to continental crust or not \citep[][]{dyar2021}. The near-global VEM dataset also provides the first-ever map of surface rock types on Venus by using FeO as a proxy for identifying gradational igneous rock types \citep[][]{dyar2020}.

VERITAS will perform multiple measurements to search for current volcanic activity that include (1) cm-scale geologic deformation, (2) recent, chemically unweathered flows, (3) volcanic thermal emission, (4) topographic or surface roughness changes, (5) near-surface water vapor, and (6) comparisons to past mission data sets including Magellan radar images and Venus Express NIR spectra at 1.02~$\upmu$m. VISAR X-band data will be compared to Magellan S-band imaging after accounting for look and wavelength differences. This approach requires that new features, such as lava flows, have different radar backscatter than the pre-existing flows. This suite of analyses is key to investigating global activity because $\approx$40$\%$ of the surface consists of ‘featureless’ plains, with limited radar backscatter variation. Flows with the same backscatter as prior flows are invisible in SAR images. Flows of interest can be mapped using the VISAR DEM. Due to the high surface pressure on Venus, evidence for outgassing, if observed, would require several percent water in the magma and would thus be an extremely valuable constraint on Venus’ interior volatile content.

VERITAS will create a global inventory of geodynamic processes to understand the alternate evolutionary path of Earth’s twin through (1) examining the origin of tesserae plateaus – possible continent-like features, (2) assessing the history of volcanism and how it has shaped Venus’ young surface, (3) looking at craters and modifications subsequent to their formation, (4) characterizing possible subduction zones and the processes governing their formation, (5) looking for evidence of prior features buried by volcanism, and (6) determining the links between interior convection and surface geology. VERITAS data will also enable the estimation of elastic thickness (a proxy for thermal gradient) and density differences due to subsurface processes (e.g., rifts, small plumes) \citep[][]{borrelli2021}. VERITAS will also constrain the interior structure, including mantle viscosity, core size and state \citep[][]{cascioli2021}.

VERITAS discoveries will lay the groundwork for future Venus missions, providing complementary data to optimize the science return of probe, aerial, or lander missions. VERITAS offers synergies with DAVINCI and ENVISION in terms of complimentary dataset resolution and coverage, and, as originally planned for a 2027 launch, an extended observational baseline with modern instrumentation. As such, these synergies represent the potential for enhanced science return and a benefit to the entire community.

\subsection{EnVision}

EnVision was selected in 2021 as ESA’s 5th Medium-class mission, scheduled for a tentative launch date in the early 2030s and a start of science operations at Venus by the end of 2034 or the beginning of 2035. 
The mission is a partnership between ESA and NASA, where NASA provides the Synthetic Aperture Radar payload, VenSAR. EnVision will investigate Venus from its inner core to its atmosphere at an unprecedented scale of resolution, characterizing, in particular, the core and mantle structure, and signs of active or past geologic processes. It addresses the overarching goals of constraining the evolution of Venus by focusing on the exchanges of volatiles from the surface to the cloud-top. It will carry a robust suite of instruments including VenSAR, three spectrometers VenSpec-M, VenSpec-U and VenSpec-H designed to observe the surface and atmosphere of Venus, and SRS, a high-frequency radar sounding instrument to penetrate the subsurface. Data from this suite of instruments, coupled with gravity science based on tracking data and radio occultation measurements will support scientific investigations of the surface, interior and atmosphere and their various interactions.

EnVision will deliver new insights into the planet's geological history through complementary imagery, polarimetry, radiometry and spectroscopy of the surface coupled with subsurface sounding and gravity mapping; it will search for thermal, morphological, and gaseous signs of volcanic and other geological activity. Key volatile species (water, HDO, SO$_2$, CO, COS) will be tracked from their sources and sinks at the surface through the clouds up to the mesosphere. Following the same approach through which our understanding of Earth and Mars has been developed, EnVision will combine global observations at low or moderate spatial resolution (e.g. surface emissivity and atmosphere composition) with regionally targeted observations using higher spatial resolutions from SAR and subsurface sounding radar profiles. EnVision’s required science orbit is a low, quasi-polar orbit with an inclination between 87 and 89$^{\circ}$, an altitude of 220 to 510~km above ground, with a resulting orbital period of $\approx$ 92~min. The choice of science orbit around Venus is mostly driven by the radio science (RSE) gravity experiment and the operational altitude range for the SAR and SRS instruments. 

EnVision’s VenSAR radar, designed and built by NASA's Jet Propulsion Laboratory, will contribute to addressing some of the key science objectives of the mission. It will image pre-selected Regions of Interest with a resolution of 30~m/pixel and high-resolution images (10~m/pixel) for more than 2\% of the surface during the nominal mission. Surface emissivity and roughness will be derived from the imaging in HV (horizontal transmit, vertical receive) and HH (horizontal transmit, horizontal receive) polarizations as well as passive radiometry. Comparison to the Magellan images and with the VenSAR data set will allow searching for surface changes due to volcanic, tectonic and landscape-forming processes on a year-to-decade timescale. Used in passive radiometry mode, VenSAR will map the thermal emission emanating from Venus’ surface with significantly better precision and accuracy than the Magellan radar. Emission maps, in the form of surface brightness temperature maps, will then be used to search for thermal anomalies or, if the surface temperature is known, to map the emissivity of the surface which, in turn, provides insights into its composition (through the dielectric constant) and physical properties (roughness, density). Decadal timescale surface modifications will be assessed by comparing EnVision images to Magellan data. 

The characterization of past and/or current volcanic activity will be conducted not only by surface changes but also by monitoring atmospheric variabilities such as water vapor or volcanic ash plumes, over the course of the six-Venus sidereal days nominal mission (4 Earth years). Species that have been observed to be the most variable in the atmosphere of Venus - SO$_2$, SO, H$_2$O, CO, COS, H$_2$2SO$_4$ - are often associated with volcanic emissions on Earth. The goal of EnVision is to understand the intrinsic atmospheric variability on a range of time scales ranging from hours to years, and how it can be associated with extrinsic inputs such as complex geological terrains or volcanotectonic features. Using its complementary suite of instruments as well as the Radio Science Experiment, EnVision will search for water vapor anomalies in three different altitude bands. Measurement of gaseous species below the clouds at altitudes of 0 - 50~km will be achieved thanks to several IR spectral transparency “windows” at around 1~$\upmu$m, at 1.17~$\upmu$m and 2.3~$\upmu$m; within the clouds, by characterizing the attenuation in the clouds of thermal emission from the low atmosphere; and above the clouds, by investigating variations of sulfur and water vapor species and related cloud properties. The RSE will also, through gravimetry, provide essential data (like the tidal Love number k$_2$) regarding the structure of the planet's interior, like the size of the core, and possibly its state (liquid or solid), which will prove invaluable as constraints for long-term geodynamic evolution models.

Several key gases will be mapped below the cloud deck by VenSpec-H: water vapor (H$_2$O and HDO), sulfur compounds (SO$_2$, COS) and carbon monoxide (CO) - these are all potential volcanic gases. This will provide estimates regarding the interior composition and the redox state, and it will help to improve the modeling of outgassing processes throughout the history of Venus. In particular, discovering spatial variability of the D/H ratio – whether associated with volcanic plumes or other fractionating processes – would be fundamental for understanding the history of water on Venus. EnVision, with VenSPEC-U, will also resume the observations of trace gases in the mesosphere conducted by previous instruments such as VEx/SPICAV-UV \citep[][]{marcq2013,marcq2020} and HST/STIS \citep[][]{jessup2015}. Simultaneous and co-located IR observations will allow simultaneous retrieval of water vapor, carbon monoxide and further key tracer species, such as SO2, OCS, HDO or hydrogen halides \citep[][]{Marcq2023}. VenSpec-M micron surface emission data at $\approx$1 will constrain surface mineralogy of features like tesserae, thus allowing to search for felsic crust and/or sedimentary rocks, while also studying cloud microphysics and dynamics. Together, these observations will provide an unprecedented view of transport and chemical processes at the cloud-top, and a better understanding of volatile transport throughout the atmosphere. 

\subsection{Other advanced missions}

Other missions are being prepared at this time. VENERA-D (Roskosmos) is designed as an ambitious successor of the VENERA and Vega missions and aims at providing surface in-situ measurements and extending the duration of the presence of instruments at the surface of Venus. It is currently in phase A development stage. The preferred targeted area, for now, is the regional plains. VENERA-D features a very wide set of scientific objectives. It should include an orbiter focused on the thermal structure and composition of the atmosphere, that will also cover its dynamics, the magnetosphere, atmospheric escape, isotopic measurements, and study of the clouds. A lander is also included for descent measurements and geophysical experiments at the surface (Venera-D Joint Science Definition Team, 2019). This lander part of the proposed mission, in particular, would be able to bring unique in-situ measurements related to the local subsurface  composition and properties or the surface mineralogy, including, for example, radiogenic elements abundances, useful as constraints on Venus crustal production evolution, and detection of water in surface minerals and their redox state (for surface-atmosphere interaction). It would also provide valuable observation of the still poorly understood near-surface atmosphere (vertical profiles of the composition, including isotopes and trace elements, and of the thermal structure of the atmosphere).

Another mission with a clear set of objectives and payload, at the moment, is the Venus Orbiter Mission, also known as Shukrayaan-1. The mission will leverage the expertise of ISRO (Indian Space Research Organisation) to map the surface of Venus with a polarymetric SAR (Synthetic Aperture Radar). The list of other instruments is not yet finalized, but the goals should include a variety of geodynamical and geological topics. The signature aim of the mission is to look at the structure of the subsurface of Venus, and possibly detect buried lava flows and craters. Shukrayaan-1 aims at identifying volcanically-active areas, by IR and SAR imaging, to constrain present-day volcanic production. It would also look at the past of Venus, first by characterizing impact craters at a high resolution to estimate how much they are altered \citep[][]{Herrick2023}, which affects our understanding of the age of the surface and past volcanic activity. It will also image tesserae, as a unique feature of Venus. Shukrayaan-1 should also include an atmospheric component, first looking at the atmosphere composition, signatures of volcanism (like SO$_2$ and its cycle) and atmosphere/clouds dynamics (super-rotation, structure, albedo). The final component of the mission will address plasma physics to characterize the escape from the upper atmosphere of Venus.

The Venus Life Finder mission is a series of probes sponsored by the Massachusetts Institute of Technology \citep[i.e.][]{seager2022venus} that would target the atmosphere of Venus with the specific goal of assessing the presence of life or signs of life. It has a more specific astrobiology goal, than the more general calling of the other missions discussed above. The first small probe is proposed to launch in 2025 and is set for short-duration atmospheric entry. It will make in-situ measurements, searching for organic compounds at the cloud level and their composition and shape. It will also retrieve additional data on the pressure, altitude and temperature of the atmosphere. A second later stage of the mission will involve the deployment of a balloon in the cloud layer to make in-situ measurements (in particular, pH, O$_2$, NH$_3$, PH$_3$, H$_2$, and non-volatile elements) and estimate its habitability. A third probe could then include a sample return mission.

Finally, Japan and China have also announced orbiters proposed for
launch before the end of this decade along the same scientific guidelines as discussed above \citep[][]{antonita2022overview,widemann2023}.

\subsection{Future missions and concepts}

Future mission concepts \citep[][]{limaye2023} aim at closing the gaps that are still left after the current missions have reached Venus. As discussed before, the structure of the interior of Venus is still poorly constrained. They will also have to mitigate the harsh environment of Venus' surface and lower atmosphere. This is why balloons have long been suggested as a valid approach while staying at a relatively safe altitude \citep[i.e.][for atmosphere studies]{ainsworth1970}, for extended missions \citep[][]{bugga2022,arredondo2022}. 
Balloons allow direct access to areas of interest, such as the clouds themselves \citep[i.e.][for habitability studies]{agrawal2022,seager2023}.
Indeed, atmospheric studies (dynamics, chemistry, radiative balance) would obviously benefit from aerial platforms, and other prime targets for study include the detection of volcanic gases \citep[][]{Marcq2023} and isotopic measurements of volatile elements including noble gases \citep{avice_noble_2022}.
Balloons could also be an interesting platform for the solid planet, in particular, for airborne seismometers \citep[][]{krishnamoorthy2023floatilla} that could last long enough to measure Venus-quakes. This would constitute an alternative to the surface seismometer missions \citep[i.e.][]{kremic2020} that require high-temperature electronics to reach a goal of 120 days of operation. Airborne seismometers have been successfully tested on Earth \citep[for example][]{garcia2022} and should benefit from the high density of Venus' atmosphere \citep[][]{lognonne2015}. Such measurements would finally provide constraints on the structure and tectonics of the planet (core radius, mantle thickness, global level of seismic activity), which we are still sorely lacking even though they may prove critical for internal modeling. 
Balloons could also be a platform for magnetometers \citep[][]{arredondo2022}, in a bid to detect possible ancient remanent magnetic fields in old crustal material \citep[i.e.][]{orourke2018a}, which could help constrain the thermal history and structure of Venus.

However, it is likely that some questions could remain unanswered until in-situ measurements are made at the surface of Venus, with mineralogical sensors and instruments like drills. This would require a considerable investment in new technology in terms of high-temperature electronics \citep[][]{wilson2016venus} and improved cooling systems, considering the high-pressure, high-temperature Venusian environment. Notably, surface probes in the past (and including proposed missions) targeted Venus' plains for security. Presumably, contrasting those measurements with in-situ data from tesserae terrains would greatly improve our understanding of the relevance of these landforms in the context of Venus' history. Other uses of surface probes include the measurement of the surface heat flux, despite probable spatial variations, that informs the lithosphere thickness and constrains interior models of the structure, dynamics and composition of Venus \citep[][]{Rolf2022,Gillmann2022}. Finally, in-situ measurements could bring a trove of measurements on the thermal properties, structure and dynamics of the lower and near-surface atmosphere, which is still poorly understood at present-day.
Another type of mission could be a sample return mission, with the collection of samples of gases contained in the Venus atmosphere and their return to Earth. This type of mission presents important challenges, notably because the atmospheric samples would be collected during a hypervelocity entry ($>$10~km~s$^{-1}$, see \citet{rabinovitch_hypervelocity_2019}), involving partial ionization of the gas in the plasma in front of the probe and potential elemental and/or isotopic fractionation of the sampled gas. However, the potential scientific discoveries allowed by the precise measurements of Venus gases with state-of-the-art techniques installed in Earth-based laboratories represent an important motivation to explore the feasibility of such mission concepts \citep[][]{widemann2023}.  

\section{Conclusion}

Our understanding of Venus has advanced in leaps over the last decades despite the relatively low number of missions targeting this planet compared to Mars, for example. The scientific community now has access to a wide range of measurements pertaining to most aspects of the planet, and many key discoveries have been made. Notable recent developments include a better grasp of the atmosphere dynamics and the origin of its super-rotation, but also a growing realisation that Venus is very likely still an active world, from an internal point of view, albeit a very different one compared to Earth. Unfortunately, observations have been complicated by a relatively young surface and hellish atmospheric conditions. For these reasons, it is the atmosphere that received most of the attention lately. Comparatively, we still know very little for sure about the solid planet: interior properties remain mostly unknown and are generally assumed to be comparable to Earth's, while even the surface characteristics of Venus are poorly understood. Thankfully, we have finally reached a point where we have a good idea of what we do not know and how to get there. Additionally, the importance of Venus is now being recognized in the solar system and exoplanet studies. New selected, and proposed missions are set to close the gaps in the data. While the upcoming missions to Venus will not bring all the answers we need, they go after one of the biggest identified mysteries - the past evolution of Venus - with a multidisciplinary approach designed to use as many available constraints as possible.

\section*{Acknowledgements}
The authors thank the anonymous reviewer for his constructive comments that helped to improve the manuscript. CG acknowledges that this work has been carried out within the framework of the NCCR PlanetS supported by the Swiss National Science Foundation under grants 51NF40$\_182901$ and 51NF40$\_205606$.
ML acknowledges support by a postdoctoral grant from France's Centre National d'\'Etudes Spatiales (CNES). ML acknowledges funding from the European Union's Horizon Europe research and innovation program under the Marie Sk\l odowska-Curie grant agreement 101110489/MuSICA-V. GA acknowledges France’s Centre National d’Études Spatiales (CNES) for funding support of Venus related studies.

\bibliographystyle{elsarticle-harv} 
\bibliography{elsarticle-template-harv}





\end{document}